\documentclass[aps,prl,article,twocolumn,preprintnumbers,amsmath,amssymb,superscriptaddress]{revtex4-2}

\usepackage{graphicx}  
\usepackage{dcolumn}   
\usepackage{bm}        
\usepackage{amssymb}   
\usepackage{amsmath}
\usepackage{mathrsfs}
\usepackage{epigraph}
\usepackage{braket}
\usepackage{gensymb}
\usepackage{lmodern}
\usepackage{ tipa }
\usepackage{bbold}
\usepackage{esint}
\usepackage{mathdots}
\usepackage{xcolor}
\usepackage{appendix}
\usepackage{natbib}
\usepackage{multirow}
\usepackage{float}
\usepackage{comment}
\usepackage{xr}
\usepackage[normalem]{ulem}
\usepackage[colorlinks=true, linkcolor=blue, urlcolor=blue, citecolor=blue]{hyperref}

\begin{document}
\title{Signatures of the Attractive Interaction in Spin Spectra of One-Dimensional Cuprate Chains} 	
\author{Zecheng Shen}\thanks{These authors contributed equally to this work.}
\affiliation{Department of Chemistry, Emory University, Atlanta, Georgia 30322, USA}
\author{Jiarui Liu}\thanks{These authors contributed equally to this work.}
\affiliation{Department of Physics and Astronomy, Clemson University, Clemson, South Carolina 29634, USA}
\affiliation{Department of Physics, University of California, Berkeley, California 94720, USA}
\author{Hao-Xin Wang}
\affiliation{Department of Physics, The Chinese University of Hong Kong, Shatin, New Territories 999077, Hong Kong, China}
\author{Yao Wang}
\email[\href{mailto:yao.wang@emory.edu}{yao.wang@emory.edu}]{}
\affiliation{Department of Chemistry, Emory University, Atlanta, Georgia 30322, USA}
\affiliation{Department of Physics and Astronomy, Clemson University, Clemson, South Carolina 29634, USA}

\date{\today}
\begin{abstract}
    Identifying the minimal model for cuprates is crucial for explaining the high-$T_c$ pairing mechanism. Recent photoemission experiments have suggested a significant near-neighbor attractive interaction $V$ in cuprate chains, favoring pairing instability. To determine its strength, we systematically investigate the dynamical spin structure factors $S(q,\omega)$ using the density matrix renormalization group. Our analysis quantitatively reveals a notable softening in the two-spinon continuum, particularly evident in the intense spectrum at large momentum. This softening is primarily driven by the renormalization of the superexchange interaction, as determined by a comparison with the slave-boson theory. We also demonstrate the feasibility of detecting this spectral shift in thin-film samples using resonant inelastic x-ray scattering. Therefore, this provides a distinctive fingerprint for the attractive interaction, motivating future experiments to unveil essential ingredients in cuprates.    
\end{abstract}

\maketitle

Understanding the pairing mechanism in high-$T_c$ superconductors, particularly in cuprates, is essential for technological advancements in energy and quantum information science. It also presents a significant challenge to the solid-state theory, driving the studies on quantum materials\,\cite{keimer2017physics} and the development of various quantum many-body numerical methods\,\cite{leblanc2015solutions, motta2017towards, qin2022hubbard}. The prevailing view is that a $d$-wave pairing instability arises from strong electronic correlations\,\cite{keimer2015quantum, scalapino2012common}, describable by the Hubbard model. With the advancements in numerical techniques, represented by the density matrix renormalization group (DMRG) and quantum Monte Carlo, recent discussions have progressed towards an unbiased evaluation of superconductivity and other emergent phases in models akin to Hubbard. In this context, the single-band Hubbard model has successfully produced multiple phases in cuprates, including the antiferromagnetism\,\cite{hirsch1985two, polatsek1996ground}, stripe fluctuations\,\cite{zheng2017stripe, huang2017numerical, ponsioen2019period}, and strange metallicity\,\cite{kokalj2017bad, huang2019strange, cha2020slope}. Although the Hubbard model and its derived $t$-$J$ model have demonstrated quasi-long-range superconductivity with specific geometries and parameters\,\cite{jiang2019superconductivity, jiang2022stripe, jiang2020ground, jiang2024ground}, determining whether these models can sustain robust superconductivity in the two-dimensional (2D) thermodynamic limit, especially on the hole-doped side, remains an active area of research\,\cite{qin2020absence, chung2020plaquette, jiang2021ground, gong2021robust, wietek2021stripes, jiang2022pairing, haldane2023d, jiang2023density, lu2024emergent, xu2024coexistence}. 

In parallel to expanding numerical limits, experiments have unveiled additional ingredients in cuprates beyond the Hubbard model. Polaronic features indicative of electron-phonon coupling have been observed in angle-resolved photoemission spectroscopy (ARPES)\,\cite{lanzara2001evidence, cuk2004coupling,he2018rapid, zhou2005multiple} and inelastic neutron scattering (INS)\,\cite{reznik2006electron}. More recently, ARPES studies have uncovered a pronounced holon folding feature in doped 1D cuprate chains, suggesting an additional near-neighbor attractive interaction between electrons\,\cite{chen2021anomalously}. This attraction is likely mediated by phonons\,\cite{wang2021phonon,tang2023traces,wang2024spectral}. Introducing this previously overlooked interaction into the Hubbard model significantly enhances the instability of superconductivity\,\cite{qu2022spin,jiang2022enhancing,zhang2022enhancement, peng2023enhanced, zhou2023robust}. Hence, accurately determining the form and strength of attractive interactions is crucial for the theoretical reproduction of superconductivity, potentially unraveling the enigma of the high-$T_c$ pairing mechanism.

As the only evidence for the near-neighbor attractive interaction $V$, the holon-folding peak observed in ARPES constitutes a nuanced detail within a continuum\,\cite{chen2021anomalously}. While the magnitude of this feature can demonstrate the existence of $V$, it introduces uncertainty regarding the interaction strength. More precise quantification of its magnitude is imperative within ARPES, necessitating the pursuit of additional spectral signatures. Therefore, this Letter explores the influence of this attractive $V$ on the dynamical spin structure factor $S(q,\omega)$, which can be measured by INS or resonant inelastic x-ray scattering (RIXS). By combining unbiased DMRG simulations and analytic slave-boson approximations, we identify a softening in the two-spinon continuum and elucidate its origin. This softening is most evident near $q\sim\pi$, where the spectral intensity is pronounced. Thus, it can be accurately quantified and remains robust even when considering the finite core-hole lifetime in RIXS, providing an efficient experimental fingerprint for precisely assessing microscopic interactions.

We focus on the extended-Hubbard model (EHM), described by the Hamiltonian
\begin{equation}
\mathcal{H} \mkern-2mu=\mkern-2mu -t \mkern-4mu\sum_{\langle ij \rangle, \sigma} \mkern-4mu(c^\dagger_{i\sigma} c_{j\sigma} + h.c.) + U \mkern-3mu\sum_i n_{i\uparrow} n_{i\downarrow} + V\mkern-16mu\sum_{\langle ij \rangle,\sigma,\sigma^\prime}\mkern-10mu n_{i\sigma} n_{j\sigma^\prime},
\end{equation} 
where $c_{i\sigma}$ ($c^\dagger_{i\sigma}$) annihilates (creates) an electron at site $i$ with spin $\sigma$ and $n_{i\sigma} = c^\dagger_{i\sigma}c_{i\sigma}$ denotes the local electron density. This model is particle-hole symmetric, hence we refer to the doping $x$ without distinguishing between hole or electron carriers. For 1D cuprate chains such as Ba$_{2-x}$Sr$_x$CuO$_{3+\delta}$, the nearest-neighbor (NN) hopping $t \simeq 0.6 \mathrm{eV}$\,\cite{chen2021anomalously} and we take a relatively modest on-site Coulomb interaction $U\simeq 6t$ to expand the range of magnetic excitations. In this Letter, we consider attractive an NN interaction ($V\leq0$), whose value was approximately identified as $-t$ in the ARPES study\,\cite{chen2021anomalously}. When $V=0$, the EHM simplifies to the Hubbard model. 

We employ DMRG to calculate the ground state $|G\rangle$ of an open-boundary EHM, which is numerically exact upon convergence\,\cite{schollwock2005density,white1992density}. The time evolution of the wavefunction after a local spin excitation is simulated using the time-dependent variational principle (TDVP)\,\cite {haegeman2011time}, resulting in the unequal-time correlation function
\begin{equation}
   \mathcal S_{ij}(t)=-i\theta(t)\braket{G|\,[\hat{S}^z_i(t), \hat{S}^z_j(0)]\,|G}\,.
\end{equation}
Here, $\hat{S}^z_i = (\hat{c}_{i\uparrow}^\dagger \hat{c}_{i\uparrow}-\hat{c}_{i\downarrow}^\dagger \hat{c}_{i\downarrow})/2$ represents the $z$-component spin operator at the $i$-th site. In this work, we keep the maximum bond dimension $D=2000$, with truncation error around $10^{-8}$. The time evolution step is set to $\delta t= 0.05t^{-1} $ and the evolution is truncated at $t_{\rm m}=40t^{-1}$. All the results presented in this Letter are obtained using an $L=48$ chain. The convergence of the calculation is verified in Supplemental Material (SM)\,\cite{supplement}.

A space-time Fourier transform applied to $\mathcal S_{ij}(t)$, by imposing translational symmetry, yields the dynamical spin structure factor $S(q,\omega)$. However, since our DMRG uses an open boundary condition, this introduces artifacts in the transform to momentum space. To better approximate the thermodynamic limit, we post process the correlation function using the cluster perturbation theory\,\cite{senechal2000spectral,senechal2002cluster,yang2016spectral}. A spin-exchange interaction $J=4t^2/(U-V)$ across the boundary is employed to perturbatively correct the boundary effects:
\begin{eqnarray}
    &&\quad S(q,\omega)=\frac1{L}\sum_{i,j}\left[\frac{\mathcal{S}(\omega)}{1-h(q)\mathcal{S}(\omega)}\right]_{ij} e^{iq(r_i-r_j)},\\
    &&h_{mn}(q)=Je^{iqL}  \delta_{m,n+L}\ \mathrm{and}\ \mathcal{S}_{ij}(\omega)\mkern-2mu=\mkern-2mu\int_0^{t_{\rm m}} \mkern-5mu \mathcal S_{ij}(t)e^{-i\omega t}dt.\nonumber
\end{eqnarray}
This correction mitigates the boundary effects and leads to continuous momentum resolution\,\cite{raum2020two, huang2022determinantal}. The Hanning window function is applied for the time transform.

\begin{figure}[!t]
\begin{center}
\includegraphics[width=8.5cm]{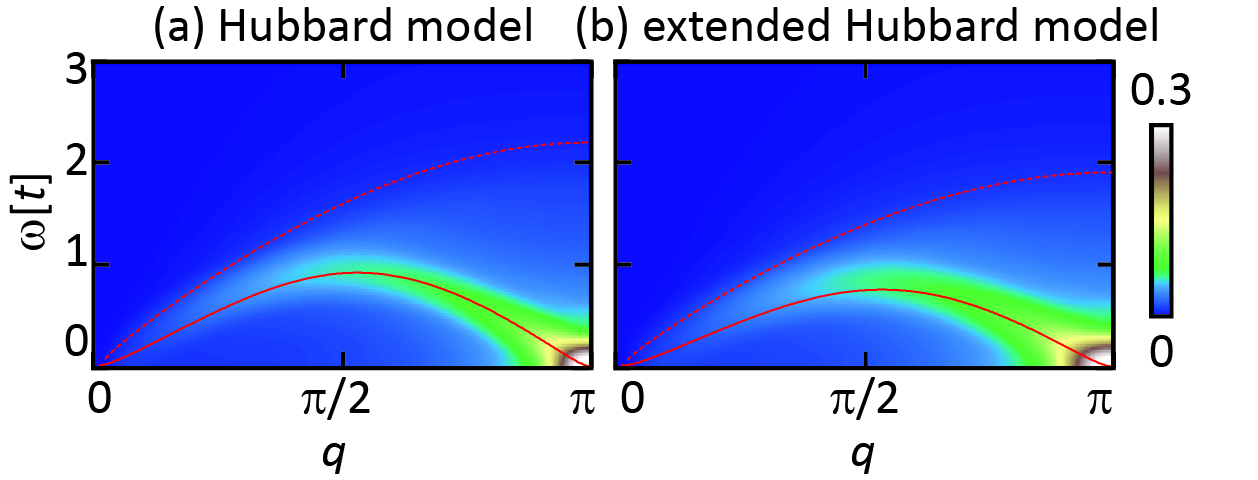}\vspace{-3mm}
\caption{Dynamical spin structure factor $S(q,\omega)$ for the half-filled (a) Hubbard model and (b) EHM with $V=-t$, calculated using DMRG in an $L=48$ chain. The red solid (dashed) line represents the lower (upper) bound derived from the Bethe ansatz for the Hubbard model and the EHM with a modified $J$.\vspace{-3mm}
}
\label{Fig:half-filled}
\end{center}
\end{figure}

In the case of the undoped parent compound, the $S(q,\omega)$ of high-quality $\text{Sr}_2\text{CuO}_{3}$ samples has been successfully analyzed through INS and RIXS\,\cite{walters2009effect, schlappa2012spin}, showing a two-spinon continuum separated from high-energy orbital excitations\,\cite{li2021particle}. As shown in Fig.~\ref{Fig:half-filled}(a), simulations based on the Hubbard model accurately reproduce this continuum. Its upper and lower bounds align with the Bethe ansatz, i.e.,~$\frac{\pi J}{2}\sin{(\frac{q}{2})}$ and $\pi J\sin{q}$\,\cite{voit1995one}. The spectral weight is predominantly concentrated near the lower bound, with a gapless excitation at $q=\pi$. Upon introducing a near-neighbor attractive interaction $V=-t$, the $S(q,\omega)$ for the EHM resembles that observed in the Hubbard model [see Fig.~\ref{Fig:half-filled}(b)]. This similarity allows for the adjustment of $J=4t^2/(U-V)$ in the Bethe ansatz, yielding upper and lower bounds that are consistent with the DMRG simulations. At $q=\pi$, where the intensity peaks, while the upper bound experiences a $14\%$ reduction, the corresponding lower bound remains gapless. Given that the spectral intensity resides primarily at the lower bound, it is challenging to characterize the strength of $V$ in the undoped system.

\begin{figure*}[htbp]
\begin{center}
\includegraphics[width=18cm]{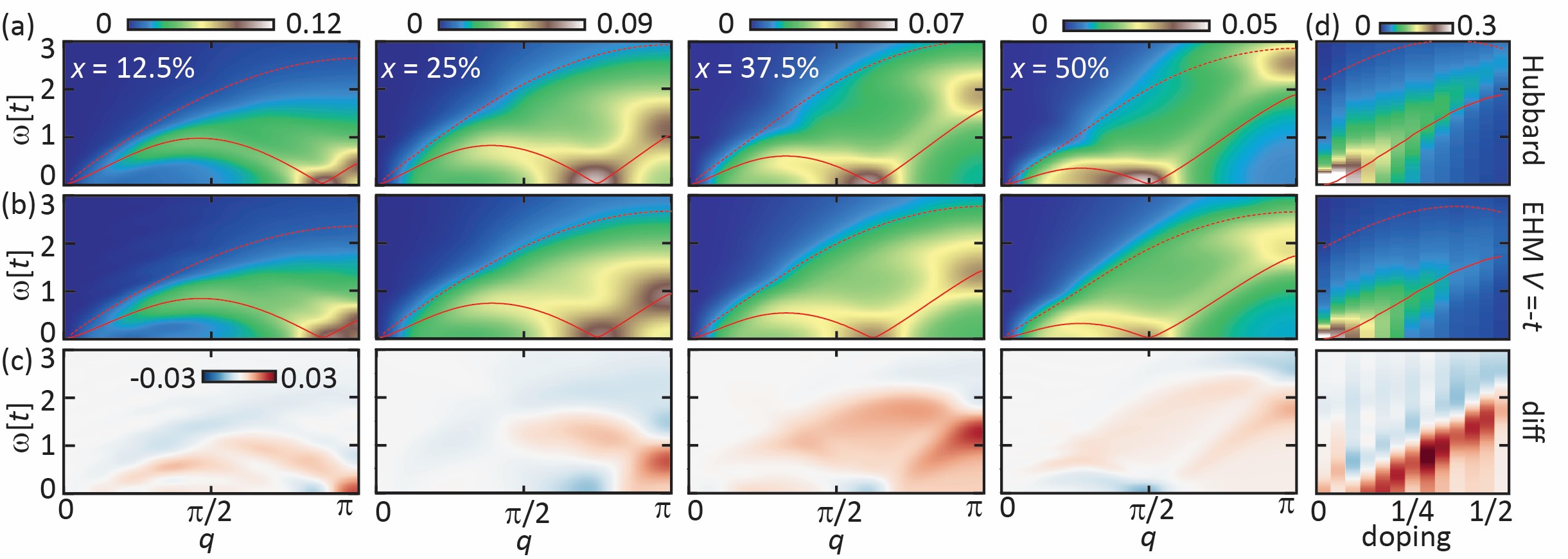}\vspace{-3mm}
\caption{The doping dependence of $S(q,\omega)$ for the (a) Hubbard model and (b) EHM with $V=-t$. The red solid (dashed) lines represent the lower (upper) bound obtained from the slave-boson mean-field theory. (c) The difference in $S(q,\omega)$ between the Hubbard model and EHM with $V=-t$, highlight the softening of two-spinon continuum. (d) A focused view on the doping dependence of the spectra at $q=\pi$ for both the Hubbard model (upper) and EHM (lower). \vspace{-3mm}
}
\label{fig:dopingDep}
\end{center}
\end{figure*}

The impact of the attractive interaction is more evident in doped systems, which can be measured through the recently synthesized Ba$_{2-x}$Sr$_x$CuO$_{3+\delta}$\,\cite{chen2021anomalously}. For the Hubbard model, doping shifts the nesting momentum $q_F$ to $(1-x)\pi$ [see Fig.~\ref{fig:dopingDep}(a)]. This shift reflects the gapless excitation between $\pm k_F$ of the spinon Fermi surface [see the SM\,\cite{supplement}]. As a result, the $q=\pi$ excitation becomes gapped. The gap size increases with doping and reaches $2.5t\sim 1.5$\,eV at 50\% doping. Unlike the undoped systems, the spectral weight maximizes at both the nesting momentum and $q=\pi$, allowing for the determination of the impact of $V$ at a finite energy. 

By incorporating the attractive $V=-t$, the $S(q,\omega)$ of the doped EHM experiences an enhancement in the momentum-integrated intensity, as the attractive $V$ suppresses doublon-hole fluctuations. In contrast, the spectral weight at the nesting momentum decreases due to the reduction in the effective superexchange $J$ [see Fig.~\ref{fig:dopingDep}(b)]. Although this nesting momentum shifts slightly toward $\pi$, it remains close to $2k_F = (1-x)\pi$, similar to the Hubbard model [see the SM\,\cite{supplement}]. Importantly, the peak energy at $q=\pi$ exhibits significant softening compared to the Hubbard model at the same doping. To better visualize this softening, we further present the differential spectra of the Hubbard and EHM model for each individual doping in Fig.~\ref{fig:dopingDep}(c). This softening increases from $0.1t$ at 12.5\% doping to $0.6t$ at 50\% doping. While the spectral intensity decreases with doping, the relatively high intensity at $q=\pi$ and the $0.5t\sim 0.3 $\,eV shift enable experimental characterization in doped 1D cuprate chains. Furthermore, as shown in Fig.~\ref{fig:dopingDep}(d), the $q=\pi$ peak energy almost linearly depends on the carrier concentration. This finding provides an additional method to extrapolate the peak positions at doping levels beyond the reach of experimental resolution.

To elucidate the reason for this softening, we estimate the bounds of spin excitations using a slave-boson mean-field approach, which has been shown to closely approximate the compact support of $S(q,\omega)$ in both Hubbard and $t-J$ models\,\cite{parschke2019numerical}. Within this framework, we first map the extended-Hubbard model into the corresponding $t$-$J$ model with $J={4t^2}/(U-V)$. The constrained electron creation operator $\Tilde{c}_{i\sigma}^\dagger$ in the $t$-$J$ model is then decomposed into $\Tilde{c}_{i\sigma}^\dagger = f_{i\sigma}^\dagger a_i,$ where $f_{i\sigma}^\dagger$ creates a spinon and $a_i$ annihilates a holon at site $i$. The exclusion of double occupancy is enforced by $a_i^\dagger a_i + \sum_\sigma f_{i\sigma}^\dagger f_{i\sigma}=1 $. By applying the resonance valence bond mean-field theory to both spinons and holons, we obtain an effective spinon dispersion $\xi_k = -(JD+2tG)\cos{k}$ [see detailed derivations in SM\,\cite{supplement}]. Its coefficients depend on the mean-field tunneling of spinons $D = \sum_\sigma \langle f_{i\sigma}^\dagger f_{i+1\sigma} \rangle= \pi \cos{({\frac{\pi x}{2})}}/2$ and holons $G = \langle a_i^\dagger a_{i+1} \rangle= \sin(\pi x)/\pi$, both of which vary with doping. Finally, $S(q,\omega)$ can be approximated by the Lindhard response function for spinons: 
\begin{equation}
    S(q,\omega)_{\rm M\mkern-1muF} = \frac{1}{4N\pi}\text{Im}\mkern-3mu\sum_k \mkern-2mu\frac{f_F(\xi_{k+q})-f_F(\xi_{k})}{\omega+\xi_k-\xi_{k+q}+i0^+}.
\end{equation}
Here, $f_F(k) = \theta(k_F-k)$ represents the Fermi-Dirac distribution at zero temperature. The upper and lower bounds derived from this slave-boson theory align with conclusions drawn using the Bethe ansatz for the half-filled Hubbard model\,\cite{parschke2019numerical}, with adjustments to $J$ in the EHM delineating the boundaries in Fig.~\ref{Fig:half-filled}(b).

\begin{figure*}[!t]
\begin{center}
\includegraphics[width=18cm]{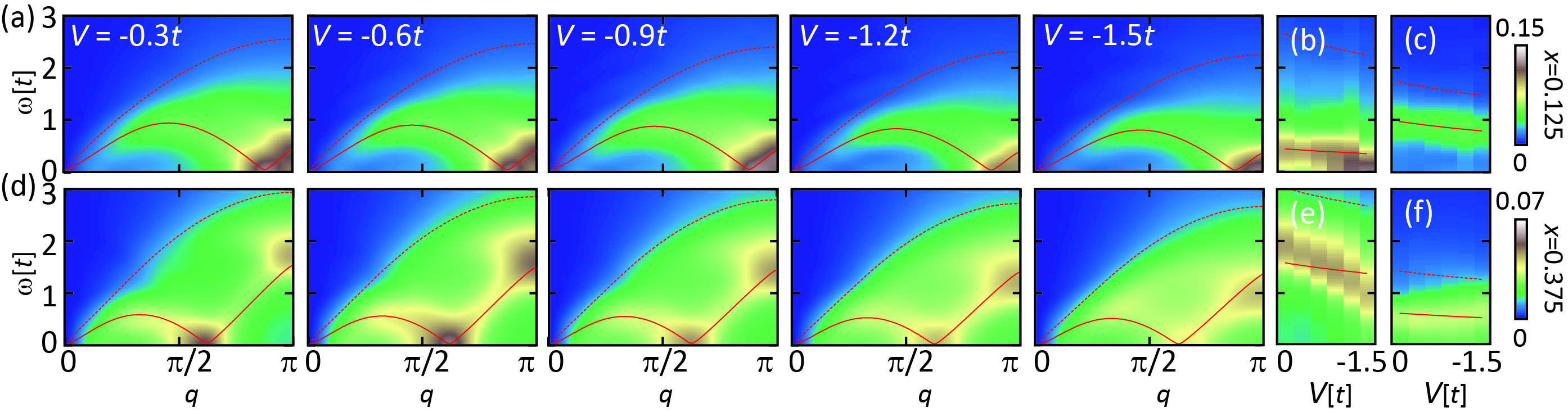}\vspace{-4mm}
\caption{(a) The impact of the near-neighbor interaction $V$ on the dynamical spin structure factor $S(q,\omega)$, with $V$ varying from $V = -0.3$ to $-1.5t$ at a doping level of $x = 12.5\%$. The red solid (dashed) lines represent the lower (upper) bound obtained from the slave-boson mean-field theory. (b),(c) The $V$ dependence of spectra at (b) $q=\pi$ and (c) $q = 7\pi/16$ (the $k_F$). (d)-(f) Same as (a)-(c) but for a higher doping level of $x = 37.5\%$. \vspace{-4mm}
}
\label{fig:VDependence}
\end{center}
\end{figure*}

At finite doping, as discussed in Figs.~\ref{fig:dopingDep}(a) and (b), the slave-boson results continue to accurately capture the upper and lower bounds of the $S(q,\omega)$ for both models below 40\% doping. This indicates that the softening of the two-spinon continuum is primarily due to the modification of the effective superexchange interaction by the attractive $V$. The lower bound starts to deviate from the simulated spectral peaks beyond 40\% doping. This deviation arises because the $t-J$ approximation of the Hubbard-like model omits correlated hoppings (commonly referred to as the three-site terms), which are negligible near half-filling but become significant at heavy doping\,\cite{parschke2019numerical, stephan1992single, jefferson1992derivation, spalek1988effect}. Given that existing experiments in Ba$_{2-x}$Sr$_x$CuO$_{3+\delta}$ achieve up to 40\% doping, the analytical estimation remains robust within the doping ranges of these accessible materials.

To establish benchmarks for accurately identifying the interactions from experiments, we explore how the $S(q,\omega)$ varies with different strengths of $V$. At 12.5\% doping [see Fig.~\ref{fig:VDependence}(a)], the spectral distribution evolves continuously from the Hubbard model ($V=0$) to EHM with $V=-1.2\,t$. Increasing $V$ further to $-1.5\,t$ triggers a phase separation\,\cite{qu2022spin}, manifested as a shift of $q_F$ to $\pi$ (see the SM\,\cite{supplement}). Apart from the phase separation, the two-spin continuum softens due to the renormalization of $J$. Its spectral distribution closely follows the slave-boson-derived bounds, particularly evident in the pronounced spin excitations at $q=\pi$ [see Fig.~\ref{fig:VDependence}(b)]. Remarkably, for 12.5\% doping, peak positions at $q=\pi$ are as low as $0.22-0.45t$, corresponding to $\sim 130-270$\,meV. To better resolve the excitation energy at low dopings, an alternative is examining the spectrum at $q=k_F$ instead, where the peak energy approaches $t$. Despite its lower intensity, the peak energy at $q=k_F$ corroborates the predictions from the slave-boson theory [see Fig.~\ref{fig:VDependence}(c)].

The dependence of $S(q,\omega)$ on varying $V$ strengths at a higher doping level (37.5\%) is shown in Fig.~\ref{fig:VDependence}(d). At this doping, the spectral distribution is significantly influenced by changes in $V$. Focusing on the pronounced spectrum at $q=\pi$ [see Fig.~\ref{fig:VDependence}(e)], its high excitation energy in the Hubbard limit allows for a more evident softening ($\sim 0.96\,t$). This linear and substantial dependence on $V$ provides a precise basis for an accurately quantifying the interaction strength experimentally. However, at such a high doping level, the deviation from the slave-boson theory and exact numerical simulations becomes more severe, especially for stronger $V$. Thus, in the context of the heavy doping regime, it is recommended to rely on simulation results rather than analytical predictions as benchmarks for future experiments.

The discussions thus far have focused on the dynamical spin structure factor $S(q,\omega)$, typically measured through INS. However, the study of doped 1D cuprate chains currently relies on thin-film Ba$_{2-x}$Sr$_x$CuO$_{3+\delta}$ samples, posing a challenge for INS experiments. Given the unavailability of single-crystal samples in the near future, an alternative to measuring $S(q,\omega)$ is the (Cu $L$-edge) RIXS. At the end of this Letter, we further discuss the feasibility of utilizing RIXS to assess the softening caused by the attractive $V$. As a two-step scattering process, the RIXS cross-section is formulated as\,\cite{ament2011resonant}
\begin{equation}
    I(\textit{\textbf{q}},\omega,\omega_{\rm in}) = \frac{1}{\pi}\text{Im}\bra{\Psi_{\rm f}}\frac{1}{\mathcal{H}-E_G-\omega-i0^+}\ket{\Psi_{\rm f}}\,,
\end{equation}
where the final-state wavefunction depends on the incident energy $\omega_{\rm in}$, in the form of
\begin{equation}
    \ket{\Psi_{\rm f}} = \sum_{j,\sigma}e^{i\textit{\textbf{q}} \cdot \textit{\textbf{r}}}_j \mathcal{D}_j^\dagger \frac{1}{\mathcal{H}'-E_G-\omega_{\rm in}-i\Gamma}\mathcal{D}_j\ket{G}\,.
\end{equation}
In these equations, $\textit{\textbf{q}}$ denotes the momentum transfer, $\omega$ the energy loss, and $E_G$ the ground-state energy. The dipole transition operator $\mathcal{D}_j^\dagger$ (at site $j$) facilitates the electronic hopping between the core and valence bands by absorbing (and emitting) an x-ray photon at a specific edge. Accordingly, the intermediate-state Hamiltonian $\mathcal{H}'$ includes a core hole and its interaction with valence electrons. Additional details regarding the RIXS simulations are provided in the SM\,\cite{supplement}. The intermediate state's lifetime is governed by the phenomenological parameter $1/\Gamma$. Previous studies have shown that RIXS, with $\pi$-$\sigma$ polarizations, effectively approximates $S(q,\omega)$ for a large $\Gamma$\,\cite{ament2009theoretical}. To examine the capability of RIXS in delineating the $V$-induced softening, we adopt $\Gamma = t$, reflecting realistic conditions in cuprates\,\cite{nyholm1981auger,rossi2019experimental}. The core-hole interaction is set to $U_c = -3t$\,\cite{tsutsui2000resonant, jia2016using}.

\begin{figure}[!b]
\begin{center}
\includegraphics[width=8.5cm]{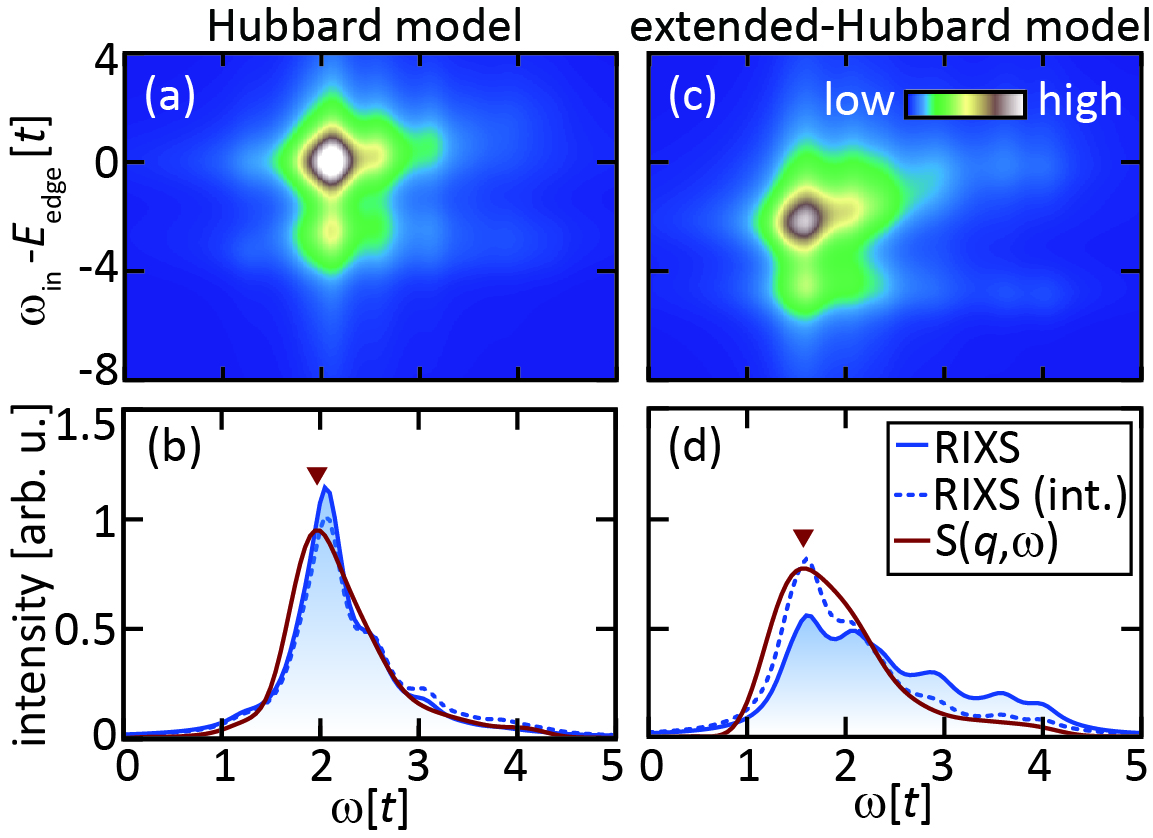}\vspace{-3mm}
\caption{(a) RIXS simulation with $q = \pi$ for a 37.5\% doped Hubbard model. (b) The comparison between RIXS at the resonant energy (blue solid line with shade), RIXS integrated along $\omega_{\rm in}$ (blue dashed line), and the $S(q,\omega)$ (red). All spectra have been normalized to the same integrated area for comparability. The red arrow highlights the peak energy in $S(q,\omega)$. (c),(d) Same as (a) and (b) but for the EHM with $V=-t$. \vspace{-5mm}}
\label{fig:RIXS}
\end{center}
\end{figure}

Figure \ref{fig:RIXS}(a) presents the RIXS spectrum for the Hubbard model at $q=\pi$, simulated using exact diagonalization on a 16-site chain. At 37.5\% doping, the spectrum exhibits two distinct resonant peaks along the incident energy ($\omega_{\rm in}$) axis, corresponding to the lower and upper Hubbard bands, respectively. Both resonances suggest the same continuum across the energy-loss axis, reflecting the two-spinon excitations. We primarily focus on the resonance at higher energy (upper Hubbard band), because its intensity exhibits a decrease with increased doping, a trend similar to that observed in the spin structure factor. As shown in Fig.~\ref{fig:RIXS}(b), the normalized RIXS cross-section at the resonant energy aligns with the $S(q,\omega)$ simulated by DMRG [the $q=\pi$ cut of Fig.~\ref{fig:dopingDep}(a)]. To reduce the influence of selecting the incident energy, we also consider the $\omega_{\rm in}$-integrated RIXS intensity, which shows minimal difference in the identified softening.

For the 37.5\% doped EHM [see Fig.~\ref{fig:RIXS}(c)], the spectral distribution experiences a softening along the incident energy $\omega_{\rm in}$ due to the reduction of effective electronic site energy with $V=-t$. At the same time, the RIXS spectrum also softens along the energy loss $\omega$ axis. A comparison between Figs.~\ref{fig:RIXS}(b) and ~\ref{fig:RIXS}(d) reveals a shift in the peak position from $2.04t$ in the Hubbard model to $1.61t$ in the EHM. 
This softening not only qualitatively mirrors the conclusion discussed above regarding $S(q,\omega)$, but also quantitatively matches the observed peak shift in $S(q,\omega)$ (from $1.95t$ to $1.54t$). The influence of the nonlocal $V$ on the intermediate state introduces more discrepancies in the spectral shape between RIXS and $S(q,\omega)$. Nonetheless, these discrepancies are mainly in the intensity distribution and do not hinder the accurate identification of the peak and lower bound of the two-spinon continuum. The SM compares RIXS and $S(q,\omega)$ at other momenta and doping\,\cite{supplement}. Hence, we find RIXS to be an efficient tool for quantifying the strength of $V$ in 1D cuprate chains.

To summarize, we have systematically investigated the spin spectra of extended-Hubbard models with various attractive interactions using DMRG. The observed softening of the two-spinon continuum, particularly pronounced near $q=\pi$, sets a numerical basis for future experiments designed to accurately quantify this interaction in cuprates. Furthermore, our discussion about RIXS spectra, with a finite core-hole lifetime, confirms that the peak softening remains precisely detectable, thereby motivating spectral characterizations on thin-film samples like Ba$_{2-x}$Sr$_x$CuO$_{3+\delta}$. This softening is primarily attributed to the renormalization of the superexchange interaction by the attractive $V$, as elucidated through comparisons with slave-boson mean-field theory. The attractive $V$ between electrons is presumably mediated by phonons, along with other longer-range interactions\,\cite{wang2021phonon}. However, these longer-range ones are not expected to alter the superexchange interaction to the lowest order and, hence not significantly impacting the spin spectra. Our choice of a relatively modest Hubbard $U$ has accounted for the phonon-mediated on-site interaction, which can be independently determined through the Mott-gap excitations measured by inelastic x-ray scattering or indirect RIXS\,\cite{abbamonte1999resonant, lu2006incident, doring2004shake}.

We thank Zhuoyu Chen, Yu He, and Wei-Sheng Lee for insightful discussions. Z.S. and Y.W. acknowledge support from the Air Force Office of Scientific Research Young Investigator Program under Grant No. FA9550-23-1-0153. This research used resources of the National Energy Research Scientific Computing Center, a DOE Office of Science User Facility supported by the Office of Science of the U.S. Department of Energy under Contract No.~DE-AC02-05CH11231 using NERSC Award No. BES-ERCAP0027096.

\bibliography{references}

\begin{thebibliography}{68}%
\makeatletter
\providecommand \@ifxundefined [1]{%
 \@ifx{#1\undefined}
}%
\providecommand \@ifnum [1]{%
 \ifnum #1\expandafter \@firstoftwo
 \else \expandafter \@secondoftwo
 \fi
}%
\providecommand \@ifx [1]{%
 \ifx #1\expandafter \@firstoftwo
 \else \expandafter \@secondoftwo
 \fi
}%
\providecommand \natexlab [1]{#1}%
\providecommand \enquote  [1]{``#1''}%
\providecommand \bibnamefont  [1]{#1}%
\providecommand \bibfnamefont [1]{#1}%
\providecommand \citenamefont [1]{#1}%
\providecommand \href@noop [0]{\@secondoftwo}%
\providecommand \href [0]{\begingroup \@sanitize@url \@href}%
\providecommand \@href[1]{\@@startlink{#1}\@@href}%
\providecommand \@@href[1]{\endgroup#1\@@endlink}%
\providecommand \@sanitize@url [0]{\catcode `\\12\catcode `\$12\catcode `\&12\catcode `\#12\catcode `\^12\catcode `\_12\catcode `\%12\relax}%
\providecommand \@@startlink[1]{}%
\providecommand \@@endlink[0]{}%
\providecommand \url  [0]{\begingroup\@sanitize@url \@url }%
\providecommand \@url [1]{\endgroup\@href {#1}{\urlprefix }}%
\providecommand \urlprefix  [0]{URL }%
\providecommand \Eprint [0]{\href }%
\providecommand \doibase [0]{https://doi.org/}%
\providecommand \selectlanguage [0]{\@gobble}%
\providecommand \bibinfo  [0]{\@secondoftwo}%
\providecommand \bibfield  [0]{\@secondoftwo}%
\providecommand \translation [1]{[#1]}%
\providecommand \BibitemOpen [0]{}%
\providecommand \bibitemStop [0]{}%
\providecommand \bibitemNoStop [0]{.\EOS\space}%
\providecommand \EOS [0]{\spacefactor3000\relax}%
\providecommand \BibitemShut  [1]{\csname bibitem#1\endcsname}%
\let\auto@bib@innerbib\@empty
\bibitem [{\citenamefont {Keimer}\ and\ \citenamefont {Moore}(2017)}]{keimer2017physics}%
  \BibitemOpen
  \bibfield  {author} {\bibinfo {author} {\bibfnamefont {B.}~\bibnamefont {Keimer}}\ and\ \bibinfo {author} {\bibfnamefont {J.}~\bibnamefont {Moore}},\ }\bibfield  {title} {\bibinfo {title} {\textit{The Physics of Quantum Materials}},\ }\href@noop {} {\bibfield  {journal} {\bibinfo  {journal} {Nat. Phys.}\ }\textbf {\bibinfo {volume} {13}},\ \bibinfo {pages} {1045} (\bibinfo {year} {2017})}\BibitemShut {NoStop}%
\bibitem [{\citenamefont {LeBlanc}\ \emph {et~al.}(2015)\citenamefont {LeBlanc}, \citenamefont {Antipov}, \citenamefont {Becca}, \citenamefont {Bulik}, \citenamefont {Chan}, \citenamefont {Chung}, \citenamefont {Deng}, \citenamefont {Ferrero}, \citenamefont {Henderson}, \citenamefont {Jim{\'e}nez-Hoyos} \emph {et~al.}}]{leblanc2015solutions}%
  \BibitemOpen
  \bibfield  {author} {\bibinfo {author} {\bibfnamefont {J.~P.}\ \bibnamefont {LeBlanc}}, \bibinfo {author} {\bibfnamefont {A.~E.}\ \bibnamefont {Antipov}}, \bibinfo {author} {\bibfnamefont {F.}~\bibnamefont {Becca}}, \bibinfo {author} {\bibfnamefont {I.~W.}\ \bibnamefont {Bulik}}, \bibinfo {author} {\bibfnamefont {G.~K.-L.}\ \bibnamefont {Chan}}, \bibinfo {author} {\bibfnamefont {C.-M.}\ \bibnamefont {Chung}}, \bibinfo {author} {\bibfnamefont {Y.}~\bibnamefont {Deng}}, \bibinfo {author} {\bibfnamefont {M.}~\bibnamefont {Ferrero}}, \bibinfo {author} {\bibfnamefont {T.~M.}\ \bibnamefont {Henderson}}, \bibinfo {author} {\bibfnamefont {C.~A.}\ \bibnamefont {Jim{\'e}nez-Hoyos}}, \emph {et~al.},\ }\bibfield  {title} {\bibinfo {title} {\textit{Solutions of the Two-Dimensional Hubbard Model: Benchmarks and Results from a Wide Range of Numerical Algorithms}},\ }\href@noop {} {\bibfield  {journal} {\bibinfo  {journal} {Phys. Rev. X}\ }\textbf {\bibinfo {volume} {5}},\ \bibinfo {pages} {041041} (\bibinfo {year}
  {2015})}\BibitemShut {NoStop}%
\bibitem [{\citenamefont {Motta}\ \emph {et~al.}(2017)\citenamefont {Motta}, \citenamefont {Ceperley}, \citenamefont {Chan}, \citenamefont {Gomez}, \citenamefont {Gull}, \citenamefont {Guo}, \citenamefont {Jim{\'e}nez-Hoyos}, \citenamefont {Lan}, \citenamefont {Li}, \citenamefont {Ma} \emph {et~al.}}]{motta2017towards}%
  \BibitemOpen
  \bibfield  {author} {\bibinfo {author} {\bibfnamefont {M.}~\bibnamefont {Motta}}, \bibinfo {author} {\bibfnamefont {D.~M.}\ \bibnamefont {Ceperley}}, \bibinfo {author} {\bibfnamefont {G.~K.-L.}\ \bibnamefont {Chan}}, \bibinfo {author} {\bibfnamefont {J.~A.}\ \bibnamefont {Gomez}}, \bibinfo {author} {\bibfnamefont {E.}~\bibnamefont {Gull}}, \bibinfo {author} {\bibfnamefont {S.}~\bibnamefont {Guo}}, \bibinfo {author} {\bibfnamefont {C.~A.}\ \bibnamefont {Jim{\'e}nez-Hoyos}}, \bibinfo {author} {\bibfnamefont {T.~N.}\ \bibnamefont {Lan}}, \bibinfo {author} {\bibfnamefont {J.}~\bibnamefont {Li}}, \bibinfo {author} {\bibfnamefont {F.}~\bibnamefont {Ma}}, \emph {et~al.},\ }\bibfield  {title} {\bibinfo {title} {\textit{Towards the Solution of the Many-Electron Problem in Real Materials: Equation of State of the Hydrogen Chain with State-of-the-Art Many-Body Methods}},\ }\href@noop {} {\bibfield  {journal} {\bibinfo  {journal} {Phys. Rev. X}\ }\textbf {\bibinfo {volume} {7}},\ \bibinfo {pages} {031059} (\bibinfo
  {year} {2017})}\BibitemShut {NoStop}%
\bibitem [{\citenamefont {Qin}\ \emph {et~al.}(2022)\citenamefont {Qin}, \citenamefont {Sch{\"a}fer}, \citenamefont {Andergassen}, \citenamefont {Corboz},\ and\ \citenamefont {Gull}}]{qin2022hubbard}%
  \BibitemOpen
  \bibfield  {author} {\bibinfo {author} {\bibfnamefont {M.}~\bibnamefont {Qin}}, \bibinfo {author} {\bibfnamefont {T.}~\bibnamefont {Sch{\"a}fer}}, \bibinfo {author} {\bibfnamefont {S.}~\bibnamefont {Andergassen}}, \bibinfo {author} {\bibfnamefont {P.}~\bibnamefont {Corboz}},\ and\ \bibinfo {author} {\bibfnamefont {E.}~\bibnamefont {Gull}},\ }\bibfield  {title} {\bibinfo {title} {\textit{The Hubbard model: A Computational Perspective}},\ }\href@noop {} {\bibfield  {journal} {\bibinfo  {journal} {Annu. Rev. Condens. Matter Phys.}\ }\textbf {\bibinfo {volume} {13}},\ \bibinfo {pages} {275} (\bibinfo {year} {2022})}\BibitemShut {NoStop}%
\bibitem [{\citenamefont {Keimer}\ \emph {et~al.}(2015)\citenamefont {Keimer}, \citenamefont {Kivelson}, \citenamefont {Norman}, \citenamefont {Uchida},\ and\ \citenamefont {Zaanen}}]{keimer2015quantum}%
  \BibitemOpen
  \bibfield  {author} {\bibinfo {author} {\bibfnamefont {B.}~\bibnamefont {Keimer}}, \bibinfo {author} {\bibfnamefont {S.~A.}\ \bibnamefont {Kivelson}}, \bibinfo {author} {\bibfnamefont {M.~R.}\ \bibnamefont {Norman}}, \bibinfo {author} {\bibfnamefont {S.}~\bibnamefont {Uchida}},\ and\ \bibinfo {author} {\bibfnamefont {J.}~\bibnamefont {Zaanen}},\ }\bibfield  {title} {\bibinfo {title} {\textit{From Quantum Matter to High-Temperature Superconductivity in Copper Oxides}},\ }\href@noop {} {\bibfield  {journal} {\bibinfo  {journal} {Nature}\ }\textbf {\bibinfo {volume} {518}},\ \bibinfo {pages} {179} (\bibinfo {year} {2015})}\BibitemShut {NoStop}%
\bibitem [{\citenamefont {Scalapino}(2012)}]{scalapino2012common}%
  \BibitemOpen
  \bibfield  {author} {\bibinfo {author} {\bibfnamefont {D.~J.}\ \bibnamefont {Scalapino}},\ }\bibfield  {title} {\bibinfo {title} {\textit{A Common Thread: The Pairing Interaction for Unconventional Superconductors}},\ }\href@noop {} {\bibfield  {journal} {\bibinfo  {journal} {Rev. Mod. Phys.}\ }\textbf {\bibinfo {volume} {84}},\ \bibinfo {pages} {1383} (\bibinfo {year} {2012})}\BibitemShut {NoStop}%
\bibitem [{\citenamefont {Hirsch}(1985)}]{hirsch1985two}%
  \BibitemOpen
  \bibfield  {author} {\bibinfo {author} {\bibfnamefont {J.~E.}\ \bibnamefont {Hirsch}},\ }\bibfield  {title} {\bibinfo {title} {\textit{Two-Dimensional Hubbard Model: Numerical Simulation Study}},\ }\href@noop {} {\bibfield  {journal} {\bibinfo  {journal} {Phys. Rev. B}\ }\textbf {\bibinfo {volume} {31}},\ \bibinfo {pages} {4403} (\bibinfo {year} {1985})}\BibitemShut {NoStop}%
\bibitem [{\citenamefont {Polatsek}\ and\ \citenamefont {Becker}(1996)}]{polatsek1996ground}%
  \BibitemOpen
  \bibfield  {author} {\bibinfo {author} {\bibfnamefont {G.}~\bibnamefont {Polatsek}}\ and\ \bibinfo {author} {\bibfnamefont {K.~W.}\ \bibnamefont {Becker}},\ }\bibfield  {title} {\bibinfo {title} {\textit{Ground-State Energy of the Hubbard Model at Half Filling}},\ }\href@noop {} {\bibfield  {journal} {\bibinfo  {journal} {Phys. Rev. B}\ }\textbf {\bibinfo {volume} {54}},\ \bibinfo {pages} {1637} (\bibinfo {year} {1996})}\BibitemShut {NoStop}%
\bibitem [{\citenamefont {Zheng}\ \emph {et~al.}(2017)\citenamefont {Zheng}, \citenamefont {Chung}, \citenamefont {Corboz}, \citenamefont {Ehlers}, \citenamefont {Qin}, \citenamefont {Noack}, \citenamefont {Shi}, \citenamefont {White}, \citenamefont {Zhang},\ and\ \citenamefont {Chan}}]{zheng2017stripe}%
  \BibitemOpen
  \bibfield  {author} {\bibinfo {author} {\bibfnamefont {B.-X.}\ \bibnamefont {Zheng}}, \bibinfo {author} {\bibfnamefont {C.-M.}\ \bibnamefont {Chung}}, \bibinfo {author} {\bibfnamefont {P.}~\bibnamefont {Corboz}}, \bibinfo {author} {\bibfnamefont {G.}~\bibnamefont {Ehlers}}, \bibinfo {author} {\bibfnamefont {M.-P.}\ \bibnamefont {Qin}}, \bibinfo {author} {\bibfnamefont {R.~M.}\ \bibnamefont {Noack}}, \bibinfo {author} {\bibfnamefont {H.}~\bibnamefont {Shi}}, \bibinfo {author} {\bibfnamefont {S.~R.}\ \bibnamefont {White}}, \bibinfo {author} {\bibfnamefont {S.}~\bibnamefont {Zhang}},\ and\ \bibinfo {author} {\bibfnamefont {G.~K.-L.}\ \bibnamefont {Chan}},\ }\bibfield  {title} {\bibinfo {title} {\textit{Stripe Order in the Underdoped Region of the Two-Dimensional Hubbard Model}},\ }\href@noop {} {\bibfield  {journal} {\bibinfo  {journal} {Science}\ }\textbf {\bibinfo {volume} {358}},\ \bibinfo {pages} {1155} (\bibinfo {year} {2017})}\BibitemShut {NoStop}%
\bibitem [{\citenamefont {Huang}\ \emph {et~al.}(2017)\citenamefont {Huang}, \citenamefont {Mendl}, \citenamefont {Liu}, \citenamefont {Johnston}, \citenamefont {Jiang}, \citenamefont {Moritz},\ and\ \citenamefont {Devereaux}}]{huang2017numerical}%
  \BibitemOpen
  \bibfield  {author} {\bibinfo {author} {\bibfnamefont {E.~W.}\ \bibnamefont {Huang}}, \bibinfo {author} {\bibfnamefont {C.~B.}\ \bibnamefont {Mendl}}, \bibinfo {author} {\bibfnamefont {S.}~\bibnamefont {Liu}}, \bibinfo {author} {\bibfnamefont {S.}~\bibnamefont {Johnston}}, \bibinfo {author} {\bibfnamefont {H.-C.}\ \bibnamefont {Jiang}}, \bibinfo {author} {\bibfnamefont {B.}~\bibnamefont {Moritz}},\ and\ \bibinfo {author} {\bibfnamefont {T.~P.}\ \bibnamefont {Devereaux}},\ }\bibfield  {title} {\bibinfo {title} {\textit{Numerical Evidence of Fluctuating Stripes in the Normal State of High-$T_{\rm c}$ Cuprate Superconductors}},\ }\href@noop {} {\bibfield  {journal} {\bibinfo  {journal} {Science}\ }\textbf {\bibinfo {volume} {358}},\ \bibinfo {pages} {1161} (\bibinfo {year} {2017})}\BibitemShut {NoStop}%
\bibitem [{\citenamefont {Ponsioen}\ \emph {et~al.}(2019)\citenamefont {Ponsioen}, \citenamefont {Chung},\ and\ \citenamefont {Corboz}}]{ponsioen2019period}%
  \BibitemOpen
  \bibfield  {author} {\bibinfo {author} {\bibfnamefont {B.}~\bibnamefont {Ponsioen}}, \bibinfo {author} {\bibfnamefont {S.~S.}\ \bibnamefont {Chung}},\ and\ \bibinfo {author} {\bibfnamefont {P.}~\bibnamefont {Corboz}},\ }\bibfield  {title} {\bibinfo {title} {\textit{Period 4 Stripe in the Extended Two-Dimensional Hubbard Model}},\ }\href@noop {} {\bibfield  {journal} {\bibinfo  {journal} {Phys. Rev. B}\ }\textbf {\bibinfo {volume} {100}},\ \bibinfo {pages} {195141} (\bibinfo {year} {2019})}\BibitemShut {NoStop}%
\bibitem [{\citenamefont {Kokalj}(2017)}]{kokalj2017bad}%
  \BibitemOpen
  \bibfield  {author} {\bibinfo {author} {\bibfnamefont {J.}~\bibnamefont {Kokalj}},\ }\bibfield  {title} {\bibinfo {title} {\textit{Bad-Metallic Behavior of Doped Mott Insulators}},\ }\href@noop {} {\bibfield  {journal} {\bibinfo  {journal} {Phys. Rev. B}\ }\textbf {\bibinfo {volume} {95}},\ \bibinfo {pages} {041110} (\bibinfo {year} {2017})}\BibitemShut {NoStop}%
\bibitem [{\citenamefont {Huang}\ \emph {et~al.}(2019)\citenamefont {Huang}, \citenamefont {Sheppard}, \citenamefont {Moritz},\ and\ \citenamefont {Devereaux}}]{huang2019strange}%
  \BibitemOpen
  \bibfield  {author} {\bibinfo {author} {\bibfnamefont {E.~W.}\ \bibnamefont {Huang}}, \bibinfo {author} {\bibfnamefont {R.}~\bibnamefont {Sheppard}}, \bibinfo {author} {\bibfnamefont {B.}~\bibnamefont {Moritz}},\ and\ \bibinfo {author} {\bibfnamefont {T.~P.}\ \bibnamefont {Devereaux}},\ }\bibfield  {title} {\bibinfo {title} {\textit{Strange Metallicity in the Doped Hubbard Model}},\ }\href@noop {} {\bibfield  {journal} {\bibinfo  {journal} {Science}\ }\textbf {\bibinfo {volume} {366}},\ \bibinfo {pages} {987} (\bibinfo {year} {2019})}\BibitemShut {NoStop}%
\bibitem [{\citenamefont {Cha}\ \emph {et~al.}(2020)\citenamefont {Cha}, \citenamefont {Patel}, \citenamefont {Gull},\ and\ \citenamefont {Kim}}]{cha2020slope}%
  \BibitemOpen
  \bibfield  {author} {\bibinfo {author} {\bibfnamefont {P.}~\bibnamefont {Cha}}, \bibinfo {author} {\bibfnamefont {A.~A.}\ \bibnamefont {Patel}}, \bibinfo {author} {\bibfnamefont {E.}~\bibnamefont {Gull}},\ and\ \bibinfo {author} {\bibfnamefont {E.-A.}\ \bibnamefont {Kim}},\ }\bibfield  {title} {\bibinfo {title} {\textit{Slope invariant T-linear Resistivity from Local Self-Energy}},\ }\href@noop {} {\bibfield  {journal} {\bibinfo  {journal} {Phys. Rev. Res.}\ }\textbf {\bibinfo {volume} {2}},\ \bibinfo {pages} {033434} (\bibinfo {year} {2020})}\BibitemShut {NoStop}%
\bibitem [{\citenamefont {Jiang}\ and\ \citenamefont {Devereaux}(2019)}]{jiang2019superconductivity}%
  \BibitemOpen
  \bibfield  {author} {\bibinfo {author} {\bibfnamefont {H.-C.}\ \bibnamefont {Jiang}}\ and\ \bibinfo {author} {\bibfnamefont {T.~P.}\ \bibnamefont {Devereaux}},\ }\bibfield  {title} {\bibinfo {title} {\textit{Superconductivity in the Doped Hubbard Model and its Interplay with Next-Nearest Hopping t'}},\ }\href@noop {} {\bibfield  {journal} {\bibinfo  {journal} {Science}\ }\textbf {\bibinfo {volume} {365}},\ \bibinfo {pages} {1424} (\bibinfo {year} {2019})}\BibitemShut {NoStop}%
\bibitem [{\citenamefont {Jiang}\ and\ \citenamefont {Kivelson}(2022)}]{jiang2022stripe}%
  \BibitemOpen
  \bibfield  {author} {\bibinfo {author} {\bibfnamefont {H.-C.}\ \bibnamefont {Jiang}}\ and\ \bibinfo {author} {\bibfnamefont {S.~A.}\ \bibnamefont {Kivelson}},\ }\bibfield  {title} {\bibinfo {title} {\textit{Stripe Order Enhanced Superconductivity in the Hubbard Model}},\ }\href@noop {} {\bibfield  {journal} {\bibinfo  {journal} {Proc. Natl. Acad. Sci.}\ }\textbf {\bibinfo {volume} {119}},\ \bibinfo {pages} {e2109406119} (\bibinfo {year} {2022})}\BibitemShut {NoStop}%
\bibitem [{\citenamefont {Jiang}\ \emph {et~al.}(2020)\citenamefont {Jiang}, \citenamefont {Zaanen}, \citenamefont {Devereaux},\ and\ \citenamefont {Jiang}}]{jiang2020ground}%
  \BibitemOpen
  \bibfield  {author} {\bibinfo {author} {\bibfnamefont {Y.-F.}\ \bibnamefont {Jiang}}, \bibinfo {author} {\bibfnamefont {J.}~\bibnamefont {Zaanen}}, \bibinfo {author} {\bibfnamefont {T.~P.}\ \bibnamefont {Devereaux}},\ and\ \bibinfo {author} {\bibfnamefont {H.-C.}\ \bibnamefont {Jiang}},\ }\bibfield  {title} {\bibinfo {title} {\textit{Ground State Phase Diagram of the Doped Hubbard Model on the Four-Leg Cylinder}},\ }\href@noop {} {\bibfield  {journal} {\bibinfo  {journal} {Phys. Rev. Res.}\ }\textbf {\bibinfo {volume} {2}},\ \bibinfo {pages} {033073} (\bibinfo {year} {2020})}\BibitemShut {NoStop}%
\bibitem [{\citenamefont {Jiang}\ \emph {et~al.}(2024)\citenamefont {Jiang}, \citenamefont {Devereaux},\ and\ \citenamefont {Jiang}}]{jiang2024ground}%
  \BibitemOpen
  \bibfield  {author} {\bibinfo {author} {\bibfnamefont {Y.-F.}\ \bibnamefont {Jiang}}, \bibinfo {author} {\bibfnamefont {T.~P.}\ \bibnamefont {Devereaux}},\ and\ \bibinfo {author} {\bibfnamefont {H.-C.}\ \bibnamefont {Jiang}},\ }\bibfield  {title} {\bibinfo {title} {\textit{Ground-State Phase Diagram and Superconductivity of the Doped Hubbard Model on Six-Leg Square Cylinders}},\ }\href@noop {} {\bibfield  {journal} {\bibinfo  {journal} {Phys. Rev. B}\ }\textbf {\bibinfo {volume} {109}},\ \bibinfo {pages} {085121} (\bibinfo {year} {2024})}\BibitemShut {NoStop}%
\bibitem [{\citenamefont {Qin}\ \emph {et~al.}(2020)\citenamefont {Qin}, \citenamefont {Chung}, \citenamefont {Shi}, \citenamefont {Vitali}, \citenamefont {Hubig}, \citenamefont {Schollw{\"o}ck}, \citenamefont {White}, \citenamefont {Zhang},\ and\ \citenamefont {on~the Many-Electron~Problem)}}]{qin2020absence}%
  \BibitemOpen
  \bibfield  {author} {\bibinfo {author} {\bibfnamefont {M.}~\bibnamefont {Qin}}, \bibinfo {author} {\bibfnamefont {C.-M.}\ \bibnamefont {Chung}}, \bibinfo {author} {\bibfnamefont {H.}~\bibnamefont {Shi}}, \bibinfo {author} {\bibfnamefont {E.}~\bibnamefont {Vitali}}, \bibinfo {author} {\bibfnamefont {C.}~\bibnamefont {Hubig}}, \bibinfo {author} {\bibfnamefont {U.}~\bibnamefont {Schollw{\"o}ck}}, \bibinfo {author} {\bibfnamefont {S.~R.}\ \bibnamefont {White}}, \bibinfo {author} {\bibfnamefont {S.}~\bibnamefont {Zhang}},\ and\ \bibinfo {author} {\bibfnamefont {S.~C.}\ \bibnamefont {on~the Many-Electron~Problem)}},\ }\bibfield  {title} {\bibinfo {title} {\textit{Absence of Superconductivity in the Pure Two-dimensional Hubbard Model}},\ }\href@noop {} {\bibfield  {journal} {\bibinfo  {journal} {Phys. Rev. X}\ }\textbf {\bibinfo {volume} {10}},\ \bibinfo {pages} {031016} (\bibinfo {year} {2020})}\BibitemShut {NoStop}%
\bibitem [{\citenamefont {Chung}\ \emph {et~al.}(2020)\citenamefont {Chung}, \citenamefont {Qin}, \citenamefont {Zhang}, \citenamefont {Schollw{\"o}ck}, \citenamefont {White} \emph {et~al.}}]{chung2020plaquette}%
  \BibitemOpen
  \bibfield  {author} {\bibinfo {author} {\bibfnamefont {C.-M.}\ \bibnamefont {Chung}}, \bibinfo {author} {\bibfnamefont {M.}~\bibnamefont {Qin}}, \bibinfo {author} {\bibfnamefont {S.}~\bibnamefont {Zhang}}, \bibinfo {author} {\bibfnamefont {U.}~\bibnamefont {Schollw{\"o}ck}}, \bibinfo {author} {\bibfnamefont {S.~R.}\ \bibnamefont {White}}, \emph {et~al.},\ }\bibfield  {title} {\bibinfo {title} {\textit{Plaquette Versus Ordinary $D$-Wave Pairing in the t'-Hubbard Model on a Width-4 Cylinder}},\ }\href@noop {} {\bibfield  {journal} {\bibinfo  {journal} {Phys. Rev. B}\ }\textbf {\bibinfo {volume} {102}},\ \bibinfo {pages} {041106} (\bibinfo {year} {2020})}\BibitemShut {NoStop}%
\bibitem [{\citenamefont {Jiang}\ \emph {et~al.}(2021)\citenamefont {Jiang}, \citenamefont {Scalapino},\ and\ \citenamefont {White}}]{jiang2021ground}%
  \BibitemOpen
  \bibfield  {author} {\bibinfo {author} {\bibfnamefont {S.}~\bibnamefont {Jiang}}, \bibinfo {author} {\bibfnamefont {D.~J.}\ \bibnamefont {Scalapino}},\ and\ \bibinfo {author} {\bibfnamefont {S.~R.}\ \bibnamefont {White}},\ }\bibfield  {title} {\bibinfo {title} {\textit{Ground-State Phase Diagram of the tt'-J Model}},\ }\href@noop {} {\bibfield  {journal} {\bibinfo  {journal} {Proc. Natl. Acad. Sci.}\ }\textbf {\bibinfo {volume} {118}},\ \bibinfo {pages} {e2109978118} (\bibinfo {year} {2021})}\BibitemShut {NoStop}%
\bibitem [{\citenamefont {Gong}\ \emph {et~al.}(2021)\citenamefont {Gong}, \citenamefont {Zhu},\ and\ \citenamefont {Sheng}}]{gong2021robust}%
  \BibitemOpen
  \bibfield  {author} {\bibinfo {author} {\bibfnamefont {S.}~\bibnamefont {Gong}}, \bibinfo {author} {\bibfnamefont {W.}~\bibnamefont {Zhu}},\ and\ \bibinfo {author} {\bibfnamefont {D.~N.}\ \bibnamefont {Sheng}},\ }\bibfield  {title} {\bibinfo {title} {\textit{Robust $D$-Wave Superconductivity in the Square-Lattice $t\text{\ensuremath{-}}J$ Model}},\ }\href@noop {} {\bibfield  {journal} {\bibinfo  {journal} {Phys. Rev. Lett.}\ }\textbf {\bibinfo {volume} {127}},\ \bibinfo {pages} {097003} (\bibinfo {year} {2021})}\BibitemShut {NoStop}%
\bibitem [{\citenamefont {Wietek}\ \emph {et~al.}(2021)\citenamefont {Wietek}, \citenamefont {He}, \citenamefont {White}, \citenamefont {Georges},\ and\ \citenamefont {Stoudenmire}}]{wietek2021stripes}%
  \BibitemOpen
  \bibfield  {author} {\bibinfo {author} {\bibfnamefont {A.}~\bibnamefont {Wietek}}, \bibinfo {author} {\bibfnamefont {Y.-Y.}\ \bibnamefont {He}}, \bibinfo {author} {\bibfnamefont {S.~R.}\ \bibnamefont {White}}, \bibinfo {author} {\bibfnamefont {A.}~\bibnamefont {Georges}},\ and\ \bibinfo {author} {\bibfnamefont {E.~M.}\ \bibnamefont {Stoudenmire}},\ }\bibfield  {title} {\bibinfo {title} {\textit{Stripes, Antiferromagnetism, and the Pseudogap in the Doped Hubbard Model at Finite Temperature}},\ }\href@noop {} {\bibfield  {journal} {\bibinfo  {journal} {Phys. Rev. X}\ }\textbf {\bibinfo {volume} {11}},\ \bibinfo {pages} {031007} (\bibinfo {year} {2021})}\BibitemShut {NoStop}%
\bibitem [{\citenamefont {Jiang}\ \emph {et~al.}(2022)\citenamefont {Jiang}, \citenamefont {Scalapino}, \citenamefont {White} \emph {et~al.}}]{jiang2022pairing}%
  \BibitemOpen
  \bibfield  {author} {\bibinfo {author} {\bibfnamefont {S.}~\bibnamefont {Jiang}}, \bibinfo {author} {\bibfnamefont {D.~J.}\ \bibnamefont {Scalapino}}, \bibinfo {author} {\bibfnamefont {S.~R.}\ \bibnamefont {White}}, \emph {et~al.},\ }\bibfield  {title} {\bibinfo {title} {\textit{Pairing Properties of the $t- t'- t''- J$ Model}},\ }\href@noop {} {\bibfield  {journal} {\bibinfo  {journal} {Phys. Rev. B}\ }\textbf {\bibinfo {volume} {106}},\ \bibinfo {pages} {174507} (\bibinfo {year} {2022})}\BibitemShut {NoStop}%
\bibitem [{\citenamefont {Chen}\ \emph {et~al.}(2023)\citenamefont {Chen}, \citenamefont {Haldane},\ and\ \citenamefont {Sheng}}]{haldane2023d}%
  \BibitemOpen
  \bibfield  {author} {\bibinfo {author} {\bibfnamefont {F.}~\bibnamefont {Chen}}, \bibinfo {author} {\bibfnamefont {F.}~\bibnamefont {Haldane}},\ and\ \bibinfo {author} {\bibfnamefont {D.}~\bibnamefont {Sheng}},\ }\bibfield  {title} {\bibinfo {title} {\textit{D-Wave and Pair-Density-Wave Superconductivity in the Square-Lattice t-J Model}},\ }\href@noop {} {\bibfield  {journal} {\bibinfo  {journal} {arXiv:2311.15092}\ } (\bibinfo {year} {2023})}\BibitemShut {NoStop}%
\bibitem [{\citenamefont {Jiang}\ \emph {et~al.}(2023)\citenamefont {Jiang}, \citenamefont {Scalapino},\ and\ \citenamefont {White}}]{jiang2023density}%
  \BibitemOpen
  \bibfield  {author} {\bibinfo {author} {\bibfnamefont {S.}~\bibnamefont {Jiang}}, \bibinfo {author} {\bibfnamefont {D.~J.}\ \bibnamefont {Scalapino}},\ and\ \bibinfo {author} {\bibfnamefont {S.~R.}\ \bibnamefont {White}},\ }\bibfield  {title} {\bibinfo {title} {\textit{Density Matrix Renormalization Group Based Downfolding of the Three-Band Hubbard Model: Importance of Density-Assisted Hopping}},\ }\href@noop {} {\bibfield  {journal} {\bibinfo  {journal} {Phys. Rev. B}\ }\textbf {\bibinfo {volume} {108}},\ \bibinfo {pages} {L161111} (\bibinfo {year} {2023})}\BibitemShut {NoStop}%
\bibitem [{\citenamefont {Lu}\ \emph {et~al.}(2024)\citenamefont {Lu}, \citenamefont {Chen}, \citenamefont {Zhu}, \citenamefont {Sheng},\ and\ \citenamefont {Gong}}]{lu2024emergent}%
  \BibitemOpen
  \bibfield  {author} {\bibinfo {author} {\bibfnamefont {X.}~\bibnamefont {Lu}}, \bibinfo {author} {\bibfnamefont {F.}~\bibnamefont {Chen}}, \bibinfo {author} {\bibfnamefont {W.}~\bibnamefont {Zhu}}, \bibinfo {author} {\bibfnamefont {D.}~\bibnamefont {Sheng}},\ and\ \bibinfo {author} {\bibfnamefont {S.-S.}\ \bibnamefont {Gong}},\ }\bibfield  {title} {\bibinfo {title} {\textit{Emergent Superconductivity and Competing Charge Orders in Hole-Doped Square-Lattice t- J Model}},\ }\href@noop {} {\bibfield  {journal} {\bibinfo  {journal} {Phys. Rev. Lett.}\ }\textbf {\bibinfo {volume} {132}},\ \bibinfo {pages} {066002} (\bibinfo {year} {2024})}\BibitemShut {NoStop}%
\bibitem [{\citenamefont {Xu}\ \emph {et~al.}(2024)\citenamefont {Xu}, \citenamefont {Chung}, \citenamefont {Qin}, \citenamefont {Schollw{\"o}ck}, \citenamefont {White},\ and\ \citenamefont {Zhang}}]{xu2024coexistence}%
  \BibitemOpen
  \bibfield  {author} {\bibinfo {author} {\bibfnamefont {H.}~\bibnamefont {Xu}}, \bibinfo {author} {\bibfnamefont {C.-M.}\ \bibnamefont {Chung}}, \bibinfo {author} {\bibfnamefont {M.}~\bibnamefont {Qin}}, \bibinfo {author} {\bibfnamefont {U.}~\bibnamefont {Schollw{\"o}ck}}, \bibinfo {author} {\bibfnamefont {S.~R.}\ \bibnamefont {White}},\ and\ \bibinfo {author} {\bibfnamefont {S.}~\bibnamefont {Zhang}},\ }\bibfield  {title} {\bibinfo {title} {\textit{Coexistence of Superconductivity with Partially Filled Stripes in the Hubbard Model}},\ }\href@noop {} {\bibfield  {journal} {\bibinfo  {journal} {Science}\ }\textbf {\bibinfo {volume} {384}},\ \bibinfo {pages} {eadh7691} (\bibinfo {year} {2024})}\BibitemShut {NoStop}%
\bibitem [{\citenamefont {Lanzara}\ \emph {et~al.}(2001)\citenamefont {Lanzara}, \citenamefont {Bogdanov}, \citenamefont {Zhou}, \citenamefont {Kellar}, \citenamefont {Feng}, \citenamefont {Lu}, \citenamefont {Yoshida}, \citenamefont {Eisaki}, \citenamefont {Fujimori}, \citenamefont {Kishio} \emph {et~al.}}]{lanzara2001evidence}%
  \BibitemOpen
  \bibfield  {author} {\bibinfo {author} {\bibfnamefont {A.}~\bibnamefont {Lanzara}}, \bibinfo {author} {\bibfnamefont {P.}~\bibnamefont {Bogdanov}}, \bibinfo {author} {\bibfnamefont {X.}~\bibnamefont {Zhou}}, \bibinfo {author} {\bibfnamefont {S.}~\bibnamefont {Kellar}}, \bibinfo {author} {\bibfnamefont {D.}~\bibnamefont {Feng}}, \bibinfo {author} {\bibfnamefont {E.}~\bibnamefont {Lu}}, \bibinfo {author} {\bibfnamefont {T.}~\bibnamefont {Yoshida}}, \bibinfo {author} {\bibfnamefont {H.}~\bibnamefont {Eisaki}}, \bibinfo {author} {\bibfnamefont {A.}~\bibnamefont {Fujimori}}, \bibinfo {author} {\bibfnamefont {K.}~\bibnamefont {Kishio}}, \emph {et~al.},\ }\bibfield  {title} {\bibinfo {title} {\textit{Evidence for Ubiquitous Strong Electron-Phonon Coupling in High-Temperature Superconductors}},\ }\href@noop {} {\bibfield  {journal} {\bibinfo  {journal} {Nature}\ }\textbf {\bibinfo {volume} {412}},\ \bibinfo {pages} {510} (\bibinfo {year} {2001})}\BibitemShut {NoStop}%
\bibitem [{\citenamefont {Cuk}\ \emph {et~al.}(2004)\citenamefont {Cuk}, \citenamefont {Baumberger}, \citenamefont {Lu}, \citenamefont {Ingle}, \citenamefont {Zhou}, \citenamefont {Eisaki}, \citenamefont {Kaneko}, \citenamefont {Hussain}, \citenamefont {Devereaux}, \citenamefont {Nagaosa} \emph {et~al.}}]{cuk2004coupling}%
  \BibitemOpen
  \bibfield  {author} {\bibinfo {author} {\bibfnamefont {T.}~\bibnamefont {Cuk}}, \bibinfo {author} {\bibfnamefont {F.}~\bibnamefont {Baumberger}}, \bibinfo {author} {\bibfnamefont {D.}~\bibnamefont {Lu}}, \bibinfo {author} {\bibfnamefont {N.}~\bibnamefont {Ingle}}, \bibinfo {author} {\bibfnamefont {X.}~\bibnamefont {Zhou}}, \bibinfo {author} {\bibfnamefont {H.}~\bibnamefont {Eisaki}}, \bibinfo {author} {\bibfnamefont {N.}~\bibnamefont {Kaneko}}, \bibinfo {author} {\bibfnamefont {Z.}~\bibnamefont {Hussain}}, \bibinfo {author} {\bibfnamefont {T.}~\bibnamefont {Devereaux}}, \bibinfo {author} {\bibfnamefont {N.}~\bibnamefont {Nagaosa}}, \emph {et~al.},\ }\bibfield  {title} {\bibinfo {title} {\textit{Coupling of the $B_{1g}$ Phonon to the Antinodal Electronic States of Bi$_2$Sr$_2$Ca$_{0.92}$Y$_{0.08}$Cu$_2$O$_{8+\delta}$}},\ }\href@noop {} {\bibfield  {journal} {\bibinfo  {journal} {Phys. Rev. Lett.}\ }\textbf {\bibinfo {volume} {93}},\ \bibinfo {pages} {117003} (\bibinfo {year} {2004})}\BibitemShut {NoStop}%
\bibitem [{\citenamefont {He}\ \emph {et~al.}(2018)\citenamefont {He}, \citenamefont {Hashimoto}, \citenamefont {Song}, \citenamefont {Chen}, \citenamefont {He}, \citenamefont {Vishik}, \citenamefont {Moritz}, \citenamefont {Lee}, \citenamefont {Nagaosa}, \citenamefont {Zaanen} \emph {et~al.}}]{he2018rapid}%
  \BibitemOpen
  \bibfield  {author} {\bibinfo {author} {\bibfnamefont {Y.}~\bibnamefont {He}}, \bibinfo {author} {\bibfnamefont {M.}~\bibnamefont {Hashimoto}}, \bibinfo {author} {\bibfnamefont {D.}~\bibnamefont {Song}}, \bibinfo {author} {\bibfnamefont {S.-D.}\ \bibnamefont {Chen}}, \bibinfo {author} {\bibfnamefont {J.}~\bibnamefont {He}}, \bibinfo {author} {\bibfnamefont {I.}~\bibnamefont {Vishik}}, \bibinfo {author} {\bibfnamefont {B.}~\bibnamefont {Moritz}}, \bibinfo {author} {\bibfnamefont {D.-H.}\ \bibnamefont {Lee}}, \bibinfo {author} {\bibfnamefont {N.}~\bibnamefont {Nagaosa}}, \bibinfo {author} {\bibfnamefont {J.}~\bibnamefont {Zaanen}}, \emph {et~al.},\ }\bibfield  {title} {\bibinfo {title} {\textit{Rapid Change of Superconductivity and Electron-Phonon Coupling through Critical Doping in Bi-2212}},\ }\href@noop {} {\bibfield  {journal} {\bibinfo  {journal} {Science}\ }\textbf {\bibinfo {volume} {362}},\ \bibinfo {pages} {62} (\bibinfo {year} {2018})}\BibitemShut {NoStop}%
\bibitem [{\citenamefont {Zhou}\ \emph {et~al.}(2005)\citenamefont {Zhou}, \citenamefont {Shi}, \citenamefont {Yoshida}, \citenamefont {Cuk}, \citenamefont {Yang}, \citenamefont {Brouet}, \citenamefont {Nakamura}, \citenamefont {Mannella}, \citenamefont {Komiya}, \citenamefont {Ando} \emph {et~al.}}]{zhou2005multiple}%
  \BibitemOpen
  \bibfield  {author} {\bibinfo {author} {\bibfnamefont {X.}~\bibnamefont {Zhou}}, \bibinfo {author} {\bibfnamefont {J.}~\bibnamefont {Shi}}, \bibinfo {author} {\bibfnamefont {T.}~\bibnamefont {Yoshida}}, \bibinfo {author} {\bibfnamefont {T.}~\bibnamefont {Cuk}}, \bibinfo {author} {\bibfnamefont {W.}~\bibnamefont {Yang}}, \bibinfo {author} {\bibfnamefont {V.}~\bibnamefont {Brouet}}, \bibinfo {author} {\bibfnamefont {J.}~\bibnamefont {Nakamura}}, \bibinfo {author} {\bibfnamefont {N.}~\bibnamefont {Mannella}}, \bibinfo {author} {\bibfnamefont {S.}~\bibnamefont {Komiya}}, \bibinfo {author} {\bibfnamefont {Y.}~\bibnamefont {Ando}}, \emph {et~al.},\ }\bibfield  {title} {\bibinfo {title} {\textit{Multiple Bosonic Mode Coupling in the Electron Self-Energy of $(La_{2- x}Sr_x) CuO_4$}},\ }\href@noop {} {\bibfield  {journal} {\bibinfo  {journal} {Phys. Rev. Lett.}\ }\textbf {\bibinfo {volume} {95}},\ \bibinfo {pages} {117001} (\bibinfo {year} {2005})}\BibitemShut {NoStop}%
\bibitem [{\citenamefont {Reznik}\ \emph {et~al.}(2006)\citenamefont {Reznik}, \citenamefont {Pintschovius}, \citenamefont {Ito}, \citenamefont {Iikubo}, \citenamefont {Sato}, \citenamefont {Goka}, \citenamefont {Fujita}, \citenamefont {Yamada}, \citenamefont {Gu},\ and\ \citenamefont {Tranquada}}]{reznik2006electron}%
  \BibitemOpen
  \bibfield  {author} {\bibinfo {author} {\bibfnamefont {D.}~\bibnamefont {Reznik}}, \bibinfo {author} {\bibfnamefont {L.}~\bibnamefont {Pintschovius}}, \bibinfo {author} {\bibfnamefont {M.}~\bibnamefont {Ito}}, \bibinfo {author} {\bibfnamefont {S.}~\bibnamefont {Iikubo}}, \bibinfo {author} {\bibfnamefont {M.}~\bibnamefont {Sato}}, \bibinfo {author} {\bibfnamefont {H.}~\bibnamefont {Goka}}, \bibinfo {author} {\bibfnamefont {M.}~\bibnamefont {Fujita}}, \bibinfo {author} {\bibfnamefont {K.}~\bibnamefont {Yamada}}, \bibinfo {author} {\bibfnamefont {G.}~\bibnamefont {Gu}},\ and\ \bibinfo {author} {\bibfnamefont {J.}~\bibnamefont {Tranquada}},\ }\bibfield  {title} {\bibinfo {title} {\textit{Electron-Phonon Coupling Reflecting Dynamic Charge Inhomogeneity in Copper Oxide Superconductors}},\ }\href@noop {} {\bibfield  {journal} {\bibinfo  {journal} {Nature}\ }\textbf {\bibinfo {volume} {440}},\ \bibinfo {pages} {1170} (\bibinfo {year} {2006})}\BibitemShut {NoStop}%
\bibitem [{\citenamefont {Chen}\ \emph {et~al.}(2021)\citenamefont {Chen}, \citenamefont {Wang}, \citenamefont {Rebec}, \citenamefont {Jia}, \citenamefont {Hashimoto}, \citenamefont {Lu}, \citenamefont {Moritz}, \citenamefont {Moore}, \citenamefont {Devereaux},\ and\ \citenamefont {Shen}}]{chen2021anomalously}%
  \BibitemOpen
  \bibfield  {author} {\bibinfo {author} {\bibfnamefont {Z.}~\bibnamefont {Chen}}, \bibinfo {author} {\bibfnamefont {Y.}~\bibnamefont {Wang}}, \bibinfo {author} {\bibfnamefont {S.~N.}\ \bibnamefont {Rebec}}, \bibinfo {author} {\bibfnamefont {T.}~\bibnamefont {Jia}}, \bibinfo {author} {\bibfnamefont {M.}~\bibnamefont {Hashimoto}}, \bibinfo {author} {\bibfnamefont {D.}~\bibnamefont {Lu}}, \bibinfo {author} {\bibfnamefont {B.}~\bibnamefont {Moritz}}, \bibinfo {author} {\bibfnamefont {R.~G.}\ \bibnamefont {Moore}}, \bibinfo {author} {\bibfnamefont {T.~P.}\ \bibnamefont {Devereaux}},\ and\ \bibinfo {author} {\bibfnamefont {Z.-X.}\ \bibnamefont {Shen}},\ }\bibfield  {title} {\bibinfo {title} {\textit{Anomalously Strong Near-neighbor Attraction In Doped 1D Cuprate Chains}},\ }\href@noop {} {\bibfield  {journal} {\bibinfo  {journal} {Science}\ }\textbf {\bibinfo {volume} {373}},\ \bibinfo {pages} {1235} (\bibinfo {year} {2021})}\BibitemShut {NoStop}%
\bibitem [{\citenamefont {Wang}\ \emph {et~al.}(2021)\citenamefont {Wang}, \citenamefont {Chen}, \citenamefont {Shi}, \citenamefont {Moritz}, \citenamefont {Shen},\ and\ \citenamefont {Devereaux}}]{wang2021phonon}%
  \BibitemOpen
  \bibfield  {author} {\bibinfo {author} {\bibfnamefont {Y.}~\bibnamefont {Wang}}, \bibinfo {author} {\bibfnamefont {Z.}~\bibnamefont {Chen}}, \bibinfo {author} {\bibfnamefont {T.}~\bibnamefont {Shi}}, \bibinfo {author} {\bibfnamefont {B.}~\bibnamefont {Moritz}}, \bibinfo {author} {\bibfnamefont {Z.-X.}\ \bibnamefont {Shen}},\ and\ \bibinfo {author} {\bibfnamefont {T.~P.}\ \bibnamefont {Devereaux}},\ }\bibfield  {title} {\bibinfo {title} {\textit{Phonon-Mediated Long-Range Attractive Interaction in One-Dimensional Cuprates}},\ }\href@noop {} {\bibfield  {journal} {\bibinfo  {journal} {Phys. Rev. Lett.}\ }\textbf {\bibinfo {volume} {127}},\ \bibinfo {pages} {197003} (\bibinfo {year} {2021})}\BibitemShut {NoStop}%
\bibitem [{\citenamefont {Tang}\ \emph {et~al.}(2023)\citenamefont {Tang}, \citenamefont {Moritz}, \citenamefont {Peng}, \citenamefont {Shen},\ and\ \citenamefont {Devereaux}}]{tang2023traces}%
  \BibitemOpen
  \bibfield  {author} {\bibinfo {author} {\bibfnamefont {T.}~\bibnamefont {Tang}}, \bibinfo {author} {\bibfnamefont {B.}~\bibnamefont {Moritz}}, \bibinfo {author} {\bibfnamefont {C.}~\bibnamefont {Peng}}, \bibinfo {author} {\bibfnamefont {Z.-X.}\ \bibnamefont {Shen}},\ and\ \bibinfo {author} {\bibfnamefont {T.~P.}\ \bibnamefont {Devereaux}},\ }\bibfield  {title} {\bibinfo {title} {\textit{Traces of Electron-Phonon Coupling in One-Dimensional Cuprates}},\ }\href@noop {} {\bibfield  {journal} {\bibinfo  {journal} {Nat. Commun.}\ }\textbf {\bibinfo {volume} {14}},\ \bibinfo {pages} {3129} (\bibinfo {year} {2023})}\BibitemShut {NoStop}%
\bibitem [{\citenamefont {Wang}\ \emph {et~al.}(2024)\citenamefont {Wang}, \citenamefont {Wu}, \citenamefont {Jiang},\ and\ \citenamefont {Yao}}]{wang2024spectral}%
  \BibitemOpen
  \bibfield  {author} {\bibinfo {author} {\bibfnamefont {H.-X.}\ \bibnamefont {Wang}}, \bibinfo {author} {\bibfnamefont {Y.-M.}\ \bibnamefont {Wu}}, \bibinfo {author} {\bibfnamefont {Y.-F.}\ \bibnamefont {Jiang}},\ and\ \bibinfo {author} {\bibfnamefont {H.}~\bibnamefont {Yao}},\ }\bibfield  {title} {\bibinfo {title} {\textit{Spectral Properties of a One-dimensional Extended Hubbard Model from Bosonization and Time-dependent Variational Principle: Applications to One-dimensional Cuprates}},\ }\href@noop {} {\bibfield  {journal} {\bibinfo  {journal} {Phys. Rev. B}\ }\textbf {\bibinfo {volume} {109}},\ \bibinfo {pages} {045102} (\bibinfo {year} {2024})}\BibitemShut {NoStop}%
\bibitem [{\citenamefont {Qu}\ \emph {et~al.}(2022)\citenamefont {Qu}, \citenamefont {Chen}, \citenamefont {Jiang}, \citenamefont {Wang},\ and\ \citenamefont {Li}}]{qu2022spin}%
  \BibitemOpen
  \bibfield  {author} {\bibinfo {author} {\bibfnamefont {D.-W.}\ \bibnamefont {Qu}}, \bibinfo {author} {\bibfnamefont {B.-B.}\ \bibnamefont {Chen}}, \bibinfo {author} {\bibfnamefont {H.-C.}\ \bibnamefont {Jiang}}, \bibinfo {author} {\bibfnamefont {Y.}~\bibnamefont {Wang}},\ and\ \bibinfo {author} {\bibfnamefont {W.}~\bibnamefont {Li}},\ }\bibfield  {title} {\bibinfo {title} {\textit{Spin-triplet Pairing Induced by Near-Neighbor Attraction in the Extended Hubbard Model for Cuprate Chain}},\ }\href@noop {} {\bibfield  {journal} {\bibinfo  {journal} {Commun. Phys.}\ }\textbf {\bibinfo {volume} {5}},\ \bibinfo {pages} {257} (\bibinfo {year} {2022})}\BibitemShut {NoStop}%
\bibitem [{\citenamefont {Jiang}(2022)}]{jiang2022enhancing}%
  \BibitemOpen
  \bibfield  {author} {\bibinfo {author} {\bibfnamefont {M.}~\bibnamefont {Jiang}},\ }\bibfield  {title} {\bibinfo {title} {\textit{Enhancing d-wave Superconductivity with Nearest-neighbor Attraction in the Extended Hubbard Model}},\ }\href@noop {} {\bibfield  {journal} {\bibinfo  {journal} {Phys. Rev. B}\ }\textbf {\bibinfo {volume} {105}},\ \bibinfo {pages} {024510} (\bibinfo {year} {2022})}\BibitemShut {NoStop}%
\bibitem [{\citenamefont {Zhang}\ \emph {et~al.}(2022)\citenamefont {Zhang}, \citenamefont {Guo}, \citenamefont {Mou}, \citenamefont {Chen},\ and\ \citenamefont {Ma}}]{zhang2022enhancement}%
  \BibitemOpen
  \bibfield  {author} {\bibinfo {author} {\bibfnamefont {L.}~\bibnamefont {Zhang}}, \bibinfo {author} {\bibfnamefont {T.}~\bibnamefont {Guo}}, \bibinfo {author} {\bibfnamefont {Y.}~\bibnamefont {Mou}}, \bibinfo {author} {\bibfnamefont {Q.}~\bibnamefont {Chen}},\ and\ \bibinfo {author} {\bibfnamefont {T.}~\bibnamefont {Ma}},\ }\bibfield  {title} {\bibinfo {title} {\textit{Enhancement of $d$-Wave Pairing in the Striped Phase with Nearest Neighbor Attraction}},\ }\href@noop {} {\bibfield  {journal} {\bibinfo  {journal} {Phys. Rev. B}\ }\textbf {\bibinfo {volume} {105}},\ \bibinfo {pages} {155154} (\bibinfo {year} {2022})}\BibitemShut {NoStop}%
\bibitem [{\citenamefont {Peng}\ \emph {et~al.}(2023)\citenamefont {Peng}, \citenamefont {Wang}, \citenamefont {Wen}, \citenamefont {Lee}, \citenamefont {Devereaux},\ and\ \citenamefont {Jiang}}]{peng2023enhanced}%
  \BibitemOpen
  \bibfield  {author} {\bibinfo {author} {\bibfnamefont {C.}~\bibnamefont {Peng}}, \bibinfo {author} {\bibfnamefont {Y.}~\bibnamefont {Wang}}, \bibinfo {author} {\bibfnamefont {J.}~\bibnamefont {Wen}}, \bibinfo {author} {\bibfnamefont {Y.~S.}\ \bibnamefont {Lee}}, \bibinfo {author} {\bibfnamefont {T.~P.}\ \bibnamefont {Devereaux}},\ and\ \bibinfo {author} {\bibfnamefont {H.-C.}\ \bibnamefont {Jiang}},\ }\bibfield  {title} {\bibinfo {title} {\textit{Enhanced Superconductivity by Near-neighbor Attraction in the Doped Extended Hubbard Model}},\ }\href@noop {} {\bibfield  {journal} {\bibinfo  {journal} {Phys. Rev. B}\ }\textbf {\bibinfo {volume} {107}},\ \bibinfo {pages} {L201102} (\bibinfo {year} {2023})}\BibitemShut {NoStop}%
\bibitem [{\citenamefont {Zhou}\ \emph {et~al.}(2023)\citenamefont {Zhou}, \citenamefont {Ye}, \citenamefont {Luo}, \citenamefont {Zhao},\ and\ \citenamefont {Chang}}]{zhou2023robust}%
  \BibitemOpen
  \bibfield  {author} {\bibinfo {author} {\bibfnamefont {Z.}~\bibnamefont {Zhou}}, \bibinfo {author} {\bibfnamefont {W.}~\bibnamefont {Ye}}, \bibinfo {author} {\bibfnamefont {H.-G.}\ \bibnamefont {Luo}}, \bibinfo {author} {\bibfnamefont {J.}~\bibnamefont {Zhao}},\ and\ \bibinfo {author} {\bibfnamefont {J.}~\bibnamefont {Chang}},\ }\bibfield  {title} {\bibinfo {title} {\textit{Robust Superconducting Correlation Against Intersite Interactions in the Extended Two-Leg Hubbard Ladder}},\ }\href@noop {} {\bibfield  {journal} {\bibinfo  {journal} {Phys. Rev. B}\ }\textbf {\bibinfo {volume} {108}},\ \bibinfo {pages} {195136} (\bibinfo {year} {2023})}\BibitemShut {NoStop}%
\bibitem [{\citenamefont {Schollw{\"o}ck}(2005)}]{schollwock2005density}%
  \BibitemOpen
  \bibfield  {author} {\bibinfo {author} {\bibfnamefont {U.}~\bibnamefont {Schollw{\"o}ck}},\ }\bibfield  {title} {\bibinfo {title} {\textit{The Density-matrix Renormalization Group}},\ }\href@noop {} {\bibfield  {journal} {\bibinfo  {journal} {Rev. Mod. Phys.}\ }\textbf {\bibinfo {volume} {77}},\ \bibinfo {pages} {259} (\bibinfo {year} {2005})}\BibitemShut {NoStop}%
\bibitem [{\citenamefont {White}(1992)}]{white1992density}%
  \BibitemOpen
  \bibfield  {author} {\bibinfo {author} {\bibfnamefont {S.~R.}\ \bibnamefont {White}},\ }\bibfield  {title} {\bibinfo {title} {\textit{Density Matrix Formulation for Quantum Renormalization Groups}},\ }\href@noop {} {\bibfield  {journal} {\bibinfo  {journal} {Phys. Rev. Lett}\ }\textbf {\bibinfo {volume} {69}},\ \bibinfo {pages} {2863} (\bibinfo {year} {1992})}\BibitemShut {NoStop}%
\bibitem [{\citenamefont {Haegeman}\ \emph {et~al.}(2011)\citenamefont {Haegeman}, \citenamefont {Cirac}, \citenamefont {Osborne}, \citenamefont {Pi{\v{z}}orn}, \citenamefont {Verschelde},\ and\ \citenamefont {Verstraete}}]{haegeman2011time}%
  \BibitemOpen
  \bibfield  {author} {\bibinfo {author} {\bibfnamefont {J.}~\bibnamefont {Haegeman}}, \bibinfo {author} {\bibfnamefont {J.~I.}\ \bibnamefont {Cirac}}, \bibinfo {author} {\bibfnamefont {T.~J.}\ \bibnamefont {Osborne}}, \bibinfo {author} {\bibfnamefont {I.}~\bibnamefont {Pi{\v{z}}orn}}, \bibinfo {author} {\bibfnamefont {H.}~\bibnamefont {Verschelde}},\ and\ \bibinfo {author} {\bibfnamefont {F.}~\bibnamefont {Verstraete}},\ }\bibfield  {title} {\bibinfo {title} {\textit{Time-dependent Variational Principle for Quantum Lattices}},\ }\href@noop {} {\bibfield  {journal} {\bibinfo  {journal} {Phys. Rev. Lett.}\ }\textbf {\bibinfo {volume} {107}},\ \bibinfo {pages} {070601} (\bibinfo {year} {2011})}\BibitemShut {NoStop}%
\bibitem [{sup()}]{supplement}%
  \BibitemOpen
  \href@noop {} {}\bibinfo {note} {See Supplemental Material for discussions about the convergence of simulations, the nesting momentum shift, the slave-boson theory, and the detailed RIXS theory and simulations}\BibitemShut {NoStop}%
\bibitem [{\citenamefont {S{\'e}n{\'e}chal}\ \emph {et~al.}(2000)\citenamefont {S{\'e}n{\'e}chal}, \citenamefont {Perez},\ and\ \citenamefont {Pioro-Ladriere}}]{senechal2000spectral}%
  \BibitemOpen
  \bibfield  {author} {\bibinfo {author} {\bibfnamefont {D.}~\bibnamefont {S{\'e}n{\'e}chal}}, \bibinfo {author} {\bibfnamefont {D.}~\bibnamefont {Perez}},\ and\ \bibinfo {author} {\bibfnamefont {M.}~\bibnamefont {Pioro-Ladriere}},\ }\bibfield  {title} {\bibinfo {title} {\textit{Spectral Weight of the Hubbard Model through Cluster Perturbation Theory}},\ }\href@noop {} {\bibfield  {journal} {\bibinfo  {journal} {Phys. Rev. Lett.}\ }\textbf {\bibinfo {volume} {84}},\ \bibinfo {pages} {522} (\bibinfo {year} {2000})}\BibitemShut {NoStop}%
\bibitem [{\citenamefont {S{\'e}n{\'e}chal}\ \emph {et~al.}(2002)\citenamefont {S{\'e}n{\'e}chal}, \citenamefont {Perez},\ and\ \citenamefont {Plouffe}}]{senechal2002cluster}%
  \BibitemOpen
  \bibfield  {author} {\bibinfo {author} {\bibfnamefont {D.}~\bibnamefont {S{\'e}n{\'e}chal}}, \bibinfo {author} {\bibfnamefont {D.}~\bibnamefont {Perez}},\ and\ \bibinfo {author} {\bibfnamefont {D.}~\bibnamefont {Plouffe}},\ }\bibfield  {title} {\bibinfo {title} {\textit{Cluster Perturbation Theory for Hubbard Models}},\ }\href@noop {} {\bibfield  {journal} {\bibinfo  {journal} {Phys. Rev. B}\ }\textbf {\bibinfo {volume} {66}},\ \bibinfo {pages} {075129} (\bibinfo {year} {2002})}\BibitemShut {NoStop}%
\bibitem [{\citenamefont {Yang}\ and\ \citenamefont {Feiguin}(2016)}]{yang2016spectral}%
  \BibitemOpen
  \bibfield  {author} {\bibinfo {author} {\bibfnamefont {C.}~\bibnamefont {Yang}}\ and\ \bibinfo {author} {\bibfnamefont {A.~E.}\ \bibnamefont {Feiguin}},\ }\bibfield  {title} {\bibinfo {title} {\textit{Spectral Function of the Two-Dimensional Hubbard Model: A Density Matrix Renormalization Group Plus Cluster Perturbation Theory Study}},\ }\href@noop {} {\bibfield  {journal} {\bibinfo  {journal} {Phys. Rev. B}\ }\textbf {\bibinfo {volume} {93}},\ \bibinfo {pages} {081107} (\bibinfo {year} {2016})}\BibitemShut {NoStop}%
\bibitem [{\citenamefont {Raum}\ \emph {et~al.}(2020)\citenamefont {Raum}, \citenamefont {Alvarez}, \citenamefont {Maier},\ and\ \citenamefont {Scarola}}]{raum2020two}%
  \BibitemOpen
  \bibfield  {author} {\bibinfo {author} {\bibfnamefont {P.}~\bibnamefont {Raum}}, \bibinfo {author} {\bibfnamefont {G.}~\bibnamefont {Alvarez}}, \bibinfo {author} {\bibfnamefont {T.}~\bibnamefont {Maier}},\ and\ \bibinfo {author} {\bibfnamefont {V.}~\bibnamefont {Scarola}},\ }\bibfield  {title} {\bibinfo {title} {\textit{Two-Particle Correlation Functions in Cluster Perturbation Theory: Hubbard Spin Susceptibilities}},\ }\href@noop {} {\bibfield  {journal} {\bibinfo  {journal} {Phys. Rev. B}\ }\textbf {\bibinfo {volume} {101}},\ \bibinfo {pages} {075122} (\bibinfo {year} {2020})}\BibitemShut {NoStop}%
\bibitem [{\citenamefont {Huang}\ \emph {et~al.}(2022)\citenamefont {Huang}, \citenamefont {Ding}, \citenamefont {Liu},\ and\ \citenamefont {Wang}}]{huang2022determinantal}%
  \BibitemOpen
  \bibfield  {author} {\bibinfo {author} {\bibfnamefont {E.~W.}\ \bibnamefont {Huang}}, \bibinfo {author} {\bibfnamefont {S.}~\bibnamefont {Ding}}, \bibinfo {author} {\bibfnamefont {J.}~\bibnamefont {Liu}},\ and\ \bibinfo {author} {\bibfnamefont {Y.}~\bibnamefont {Wang}},\ }\bibfield  {title} {\bibinfo {title} {\textit{Determinantal Quantum Monte Carlo Solver for Cluster Perturbation Theory}},\ }\href@noop {} {\bibfield  {journal} {\bibinfo  {journal} {Phys. Rev. Research}\ }\textbf {\bibinfo {volume} {4}},\ \bibinfo {pages} {L042015} (\bibinfo {year} {2022})}\BibitemShut {NoStop}%
\bibitem [{\citenamefont {Walters}\ \emph {et~al.}(2009)\citenamefont {Walters}, \citenamefont {Perring}, \citenamefont {Caux}, \citenamefont {Savici}, \citenamefont {Gu}, \citenamefont {Lee}, \citenamefont {Ku},\ and\ \citenamefont {Zaliznyak}}]{walters2009effect}%
  \BibitemOpen
  \bibfield  {author} {\bibinfo {author} {\bibfnamefont {A.~C.}\ \bibnamefont {Walters}}, \bibinfo {author} {\bibfnamefont {T.~G.}\ \bibnamefont {Perring}}, \bibinfo {author} {\bibfnamefont {J.-S.}\ \bibnamefont {Caux}}, \bibinfo {author} {\bibfnamefont {A.~T.}\ \bibnamefont {Savici}}, \bibinfo {author} {\bibfnamefont {G.~D.}\ \bibnamefont {Gu}}, \bibinfo {author} {\bibfnamefont {C.-C.}\ \bibnamefont {Lee}}, \bibinfo {author} {\bibfnamefont {W.}~\bibnamefont {Ku}},\ and\ \bibinfo {author} {\bibfnamefont {I.~A.}\ \bibnamefont {Zaliznyak}},\ }\bibfield  {title} {\bibinfo {title} {\textit{Effect of Covalent Bonding on Magnetism and the Missing Neutron Intensity in Copper Oxide Compounds}},\ }\href@noop {} {\bibfield  {journal} {\bibinfo  {journal} {Nature Physics}\ }\textbf {\bibinfo {volume} {5}},\ \bibinfo {pages} {867} (\bibinfo {year} {2009})}\BibitemShut {NoStop}%
\bibitem [{\citenamefont {Schlappa}\ \emph {et~al.}(2012)\citenamefont {Schlappa}, \citenamefont {Wohlfeld}, \citenamefont {Zhou}, \citenamefont {Mourigal}, \citenamefont {Haverkort}, \citenamefont {Strocov}, \citenamefont {Hozoi}, \citenamefont {Monney}, \citenamefont {Nishimoto}, \citenamefont {Singh} \emph {et~al.}}]{schlappa2012spin}%
  \BibitemOpen
  \bibfield  {author} {\bibinfo {author} {\bibfnamefont {J.}~\bibnamefont {Schlappa}}, \bibinfo {author} {\bibfnamefont {K.}~\bibnamefont {Wohlfeld}}, \bibinfo {author} {\bibfnamefont {K.}~\bibnamefont {Zhou}}, \bibinfo {author} {\bibfnamefont {M.}~\bibnamefont {Mourigal}}, \bibinfo {author} {\bibfnamefont {M.}~\bibnamefont {Haverkort}}, \bibinfo {author} {\bibfnamefont {V.}~\bibnamefont {Strocov}}, \bibinfo {author} {\bibfnamefont {L.}~\bibnamefont {Hozoi}}, \bibinfo {author} {\bibfnamefont {C.}~\bibnamefont {Monney}}, \bibinfo {author} {\bibfnamefont {S.}~\bibnamefont {Nishimoto}}, \bibinfo {author} {\bibfnamefont {S.}~\bibnamefont {Singh}}, \emph {et~al.},\ }\bibfield  {title} {\bibinfo {title} {\textit{Spin-Orbital Separation in the Quasi-One-Dimensional Mott Insulator Sr$_2$CuO$_3$}},\ }\href@noop {} {\bibfield  {journal} {\bibinfo  {journal} {Nature}\ }\textbf {\bibinfo {volume} {485}},\ \bibinfo {pages} {82} (\bibinfo {year} {2012})}\BibitemShut {NoStop}%
\bibitem [{\citenamefont {Li}\ \emph {et~al.}(2021)\citenamefont {Li}, \citenamefont {Nocera}, \citenamefont {Kumar},\ and\ \citenamefont {Johnston}}]{li2021particle}%
  \BibitemOpen
  \bibfield  {author} {\bibinfo {author} {\bibfnamefont {S.}~\bibnamefont {Li}}, \bibinfo {author} {\bibfnamefont {A.}~\bibnamefont {Nocera}}, \bibinfo {author} {\bibfnamefont {U.}~\bibnamefont {Kumar}},\ and\ \bibinfo {author} {\bibfnamefont {S.}~\bibnamefont {Johnston}},\ }\bibfield  {title} {\bibinfo {title} {\textit{Particle-Hole Asymmetry in the Dynamical Spin and Charge Responses of Corner-Shared 1D Cuprates}},\ }\href@noop {} {\bibfield  {journal} {\bibinfo  {journal} {Comm. Phys.}\ }\textbf {\bibinfo {volume} {4}},\ \bibinfo {pages} {217} (\bibinfo {year} {2021})}\BibitemShut {NoStop}%
\bibitem [{\citenamefont {Voit}(1995)}]{voit1995one}%
  \BibitemOpen
  \bibfield  {author} {\bibinfo {author} {\bibfnamefont {J.}~\bibnamefont {Voit}},\ }\bibfield  {title} {\bibinfo {title} {\textit{One-Dimensional Fermi Liquids}},\ }\href@noop {} {\bibfield  {journal} {\bibinfo  {journal} {Rep. Prog. Phys.}\ }\textbf {\bibinfo {volume} {58}},\ \bibinfo {pages} {977} (\bibinfo {year} {1995})}\BibitemShut {NoStop}%
\bibitem [{\citenamefont {P{\"a}rschke}\ \emph {et~al.}(2019)\citenamefont {P{\"a}rschke}, \citenamefont {Wang}, \citenamefont {Moritz}, \citenamefont {Devereaux}, \citenamefont {Chen},\ and\ \citenamefont {Wohlfeld}}]{parschke2019numerical}%
  \BibitemOpen
  \bibfield  {author} {\bibinfo {author} {\bibfnamefont {E.~M.}\ \bibnamefont {P{\"a}rschke}}, \bibinfo {author} {\bibfnamefont {Y.}~\bibnamefont {Wang}}, \bibinfo {author} {\bibfnamefont {B.}~\bibnamefont {Moritz}}, \bibinfo {author} {\bibfnamefont {T.~P.}\ \bibnamefont {Devereaux}}, \bibinfo {author} {\bibfnamefont {C.-C.}\ \bibnamefont {Chen}},\ and\ \bibinfo {author} {\bibfnamefont {K.}~\bibnamefont {Wohlfeld}},\ }\bibfield  {title} {\bibinfo {title} {\textit{Numerical Investigation of Spin Excitations in a Doped Spin Chain}},\ }\href@noop {} {\bibfield  {journal} {\bibinfo  {journal} {Phys. Rev. B}\ }\textbf {\bibinfo {volume} {99}},\ \bibinfo {pages} {205102} (\bibinfo {year} {2019})}\BibitemShut {NoStop}%
\bibitem [{\citenamefont {Stephan}\ and\ \citenamefont {Horsch}(1992)}]{stephan1992single}%
  \BibitemOpen
  \bibfield  {author} {\bibinfo {author} {\bibfnamefont {W.}~\bibnamefont {Stephan}}\ and\ \bibinfo {author} {\bibfnamefont {P.}~\bibnamefont {Horsch}},\ }\bibfield  {title} {\bibinfo {title} {\textit{Single-Particle and Optical Excitations in Doped Mott-Hubbard Insulators}},\ }\href@noop {} {\bibfield  {journal} {\bibinfo  {journal} {Int. J. Mod. Phys. B}\ }\textbf {\bibinfo {volume} {6}},\ \bibinfo {pages} {589} (\bibinfo {year} {1992})}\BibitemShut {NoStop}%
\bibitem [{\citenamefont {Jefferson}\ \emph {et~al.}(1992)\citenamefont {Jefferson}, \citenamefont {Eskes},\ and\ \citenamefont {Feiner}}]{jefferson1992derivation}%
  \BibitemOpen
  \bibfield  {author} {\bibinfo {author} {\bibfnamefont {J.}~\bibnamefont {Jefferson}}, \bibinfo {author} {\bibfnamefont {H.}~\bibnamefont {Eskes}},\ and\ \bibinfo {author} {\bibfnamefont {L.}~\bibnamefont {Feiner}},\ }\bibfield  {title} {\bibinfo {title} {\textit{Derivation of a Single-Band Model for CuO$_2$ planes by a Cell-Perturbation method}},\ }\href@noop {} {\bibfield  {journal} {\bibinfo  {journal} {Phys. Rev. B}\ }\textbf {\bibinfo {volume} {45}},\ \bibinfo {pages} {7959} (\bibinfo {year} {1992})}\BibitemShut {NoStop}%
\bibitem [{\citenamefont {Spa{\l}ek}(1988)}]{spalek1988effect}%
  \BibitemOpen
  \bibfield  {author} {\bibinfo {author} {\bibfnamefont {J.}~\bibnamefont {Spa{\l}ek}},\ }\bibfield  {title} {\bibinfo {title} {\textit{Effect of Pair Hopping and Magnitude of Intra-Atomic Interaction on Exchange-Mediated Superconductivity}},\ }\href@noop {} {\bibfield  {journal} {\bibinfo  {journal} {Phys. Rev. B}\ }\textbf {\bibinfo {volume} {37}},\ \bibinfo {pages} {533} (\bibinfo {year} {1988})}\BibitemShut {NoStop}%
\bibitem [{\citenamefont {Ament}\ \emph {et~al.}(2011)\citenamefont {Ament}, \citenamefont {Van~Veenendaal}, \citenamefont {Devereaux}, \citenamefont {Hill},\ and\ \citenamefont {Van Den~Brink}}]{ament2011resonant}%
  \BibitemOpen
  \bibfield  {author} {\bibinfo {author} {\bibfnamefont {L.~J.}\ \bibnamefont {Ament}}, \bibinfo {author} {\bibfnamefont {M.}~\bibnamefont {Van~Veenendaal}}, \bibinfo {author} {\bibfnamefont {T.~P.}\ \bibnamefont {Devereaux}}, \bibinfo {author} {\bibfnamefont {J.~P.}\ \bibnamefont {Hill}},\ and\ \bibinfo {author} {\bibfnamefont {J.}~\bibnamefont {Van Den~Brink}},\ }\bibfield  {title} {\bibinfo {title} {\textit{Resonant Inelastic X-Ray Scattering Studies of Elementary Excitations}},\ }\href@noop {} {\bibfield  {journal} {\bibinfo  {journal} {Rev. Mod. Phys.}\ }\textbf {\bibinfo {volume} {83}},\ \bibinfo {pages} {705} (\bibinfo {year} {2011})}\BibitemShut {NoStop}%
\bibitem [{\citenamefont {Ament}\ \emph {et~al.}(2009)\citenamefont {Ament}, \citenamefont {Ghiringhelli}, \citenamefont {Sala}, \citenamefont {Braicovich},\ and\ \citenamefont {van~den Brink}}]{ament2009theoretical}%
  \BibitemOpen
  \bibfield  {author} {\bibinfo {author} {\bibfnamefont {L.~J.}\ \bibnamefont {Ament}}, \bibinfo {author} {\bibfnamefont {G.}~\bibnamefont {Ghiringhelli}}, \bibinfo {author} {\bibfnamefont {M.~M.}\ \bibnamefont {Sala}}, \bibinfo {author} {\bibfnamefont {L.}~\bibnamefont {Braicovich}},\ and\ \bibinfo {author} {\bibfnamefont {J.}~\bibnamefont {van~den Brink}},\ }\bibfield  {title} {\bibinfo {title} {\textit{Theoretical Demonstration of how the Dispersion of Magnetic Excitations in Cuprate Compounds can be Determined using Resonant Inelastic X-Ray Scattering}},\ }\href@noop {} {\bibfield  {journal} {\bibinfo  {journal} {Phys. Rev. Lett.}\ }\textbf {\bibinfo {volume} {103}},\ \bibinfo {pages} {117003} (\bibinfo {year} {2009})}\BibitemShut {NoStop}%
\bibitem [{\citenamefont {Nyholm}\ \emph {et~al.}(1981)\citenamefont {Nyholm}, \citenamefont {Martensson}, \citenamefont {Lebugle},\ and\ \citenamefont {Axelsson}}]{nyholm1981auger}%
  \BibitemOpen
  \bibfield  {author} {\bibinfo {author} {\bibfnamefont {R.}~\bibnamefont {Nyholm}}, \bibinfo {author} {\bibfnamefont {N.}~\bibnamefont {Martensson}}, \bibinfo {author} {\bibfnamefont {A.}~\bibnamefont {Lebugle}},\ and\ \bibinfo {author} {\bibfnamefont {U.}~\bibnamefont {Axelsson}},\ }\bibfield  {title} {\bibinfo {title} {\textit{Auger and Coster-Kronig Broadening Effects in the 2p and 3p Photoelectron Spectra from the Metals $^{22}$Ti-$^{30}$Zn}},\ }\href@noop {} {\bibfield  {journal} {\bibinfo  {journal} {J. Phys. F: Met. Phys.}\ }\textbf {\bibinfo {volume} {11}},\ \bibinfo {pages} {1727} (\bibinfo {year} {1981})}\BibitemShut {NoStop}%
\bibitem [{\citenamefont {Rossi}\ \emph {et~al.}(2019)\citenamefont {Rossi}, \citenamefont {Arpaia}, \citenamefont {Fumagalli}, \citenamefont {Sala}, \citenamefont {Betto}, \citenamefont {Kummer}, \citenamefont {De~Luca}, \citenamefont {van~den Brink}, \citenamefont {Salluzzo}, \citenamefont {Brookes} \emph {et~al.}}]{rossi2019experimental}%
  \BibitemOpen
  \bibfield  {author} {\bibinfo {author} {\bibfnamefont {M.}~\bibnamefont {Rossi}}, \bibinfo {author} {\bibfnamefont {R.}~\bibnamefont {Arpaia}}, \bibinfo {author} {\bibfnamefont {R.}~\bibnamefont {Fumagalli}}, \bibinfo {author} {\bibfnamefont {M.~M.}\ \bibnamefont {Sala}}, \bibinfo {author} {\bibfnamefont {D.}~\bibnamefont {Betto}}, \bibinfo {author} {\bibfnamefont {K.}~\bibnamefont {Kummer}}, \bibinfo {author} {\bibfnamefont {G.~M.}\ \bibnamefont {De~Luca}}, \bibinfo {author} {\bibfnamefont {J.}~\bibnamefont {van~den Brink}}, \bibinfo {author} {\bibfnamefont {M.}~\bibnamefont {Salluzzo}}, \bibinfo {author} {\bibfnamefont {N.~B.}\ \bibnamefont {Brookes}}, \emph {et~al.},\ }\bibfield  {title} {\bibinfo {title} {\textit{Experimental Determination of Momentum-Resolved Electron-Phonon Coupling}},\ }\href@noop {} {\bibfield  {journal} {\bibinfo  {journal} {Phys. Rev. Lett.}\ }\textbf {\bibinfo {volume} {123}},\ \bibinfo {pages} {027001} (\bibinfo {year} {2019})}\BibitemShut {NoStop}%
\bibitem [{\citenamefont {Tsutsui}\ \emph {et~al.}(2000)\citenamefont {Tsutsui}, \citenamefont {Tohyama},\ and\ \citenamefont {Maekawa}}]{tsutsui2000resonant}%
  \BibitemOpen
  \bibfield  {author} {\bibinfo {author} {\bibfnamefont {K.}~\bibnamefont {Tsutsui}}, \bibinfo {author} {\bibfnamefont {T.}~\bibnamefont {Tohyama}},\ and\ \bibinfo {author} {\bibfnamefont {S.}~\bibnamefont {Maekawa}},\ }\bibfield  {title} {\bibinfo {title} {\textit{Resonant Inelastic X-Ray Scattering in One-Dimensional Copper Oxides}},\ }\href@noop {} {\bibfield  {journal} {\bibinfo  {journal} {Phys. Rev. B}\ }\textbf {\bibinfo {volume} {61}},\ \bibinfo {pages} {7180} (\bibinfo {year} {2000})}\BibitemShut {NoStop}%
\bibitem [{\citenamefont {Jia}\ \emph {et~al.}(2016)\citenamefont {Jia}, \citenamefont {Wohlfeld}, \citenamefont {Wang}, \citenamefont {Moritz},\ and\ \citenamefont {Devereaux}}]{jia2016using}%
  \BibitemOpen
  \bibfield  {author} {\bibinfo {author} {\bibfnamefont {C.}~\bibnamefont {Jia}}, \bibinfo {author} {\bibfnamefont {K.}~\bibnamefont {Wohlfeld}}, \bibinfo {author} {\bibfnamefont {Y.}~\bibnamefont {Wang}}, \bibinfo {author} {\bibfnamefont {B.}~\bibnamefont {Moritz}},\ and\ \bibinfo {author} {\bibfnamefont {T.~P.}\ \bibnamefont {Devereaux}},\ }\bibfield  {title} {\bibinfo {title} {\textit{Using RIXS to Uncover Elementary Charge and Spin Excitations}},\ }\href@noop {} {\bibfield  {journal} {\bibinfo  {journal} {Phys. Rev. X}\ }\textbf {\bibinfo {volume} {6}},\ \bibinfo {pages} {021020} (\bibinfo {year} {2016})}\BibitemShut {NoStop}%
\bibitem [{\citenamefont {Abbamonte}\ \emph {et~al.}(1999)\citenamefont {Abbamonte}, \citenamefont {Burns}, \citenamefont {Isaacs}, \citenamefont {Platzman}, \citenamefont {Miller}, \citenamefont {Cheong},\ and\ \citenamefont {Klein}}]{abbamonte1999resonant}%
  \BibitemOpen
  \bibfield  {author} {\bibinfo {author} {\bibfnamefont {P.}~\bibnamefont {Abbamonte}}, \bibinfo {author} {\bibfnamefont {C.}~\bibnamefont {Burns}}, \bibinfo {author} {\bibfnamefont {E.}~\bibnamefont {Isaacs}}, \bibinfo {author} {\bibfnamefont {P.}~\bibnamefont {Platzman}}, \bibinfo {author} {\bibfnamefont {L.}~\bibnamefont {Miller}}, \bibinfo {author} {\bibfnamefont {S.}~\bibnamefont {Cheong}},\ and\ \bibinfo {author} {\bibfnamefont {M.}~\bibnamefont {Klein}},\ }\bibfield  {title} {\bibinfo {title} {\textit{Resonant Inelastic X-ray Scattering from Valence Excitations in Insulating Copper Oxides}},\ }\href@noop {} {\bibfield  {journal} {\bibinfo  {journal} {Phys. Rev. Lett.}\ }\textbf {\bibinfo {volume} {83}},\ \bibinfo {pages} {860} (\bibinfo {year} {1999})}\BibitemShut {NoStop}%
\bibitem [{\citenamefont {Lu}\ \emph {et~al.}(2006)\citenamefont {Lu}, \citenamefont {Hancock}, \citenamefont {Chabot-Couture}, \citenamefont {Ishii}, \citenamefont {Vajk}, \citenamefont {Yu}, \citenamefont {Mizuki}, \citenamefont {Casa}, \citenamefont {Gog},\ and\ \citenamefont {Greven}}]{lu2006incident}%
  \BibitemOpen
  \bibfield  {author} {\bibinfo {author} {\bibfnamefont {L.}~\bibnamefont {Lu}}, \bibinfo {author} {\bibfnamefont {J.}~\bibnamefont {Hancock}}, \bibinfo {author} {\bibfnamefont {G.}~\bibnamefont {Chabot-Couture}}, \bibinfo {author} {\bibfnamefont {K.}~\bibnamefont {Ishii}}, \bibinfo {author} {\bibfnamefont {O.}~\bibnamefont {Vajk}}, \bibinfo {author} {\bibfnamefont {G.}~\bibnamefont {Yu}}, \bibinfo {author} {\bibfnamefont {J.}~\bibnamefont {Mizuki}}, \bibinfo {author} {\bibfnamefont {D.}~\bibnamefont {Casa}}, \bibinfo {author} {\bibfnamefont {T.}~\bibnamefont {Gog}},\ and\ \bibinfo {author} {\bibfnamefont {M.}~\bibnamefont {Greven}},\ }\bibfield  {title} {\bibinfo {title} {\textit{Incident energy and polarization-dependent resonant inelastic x-ray Scattering Study of La${_2}$CuO$_4$}},\ }\href@noop {} {\bibfield  {journal} {\bibinfo  {journal} {Phys. Rev. B}\ }\textbf {\bibinfo {volume} {74}},\ \bibinfo {pages} {224509} (\bibinfo {year} {2006})}\BibitemShut {NoStop}%
\bibitem [{\citenamefont {D{\"o}ring}\ \emph {et~al.}(2004)\citenamefont {D{\"o}ring}, \citenamefont {Sternemann}, \citenamefont {Kaprolat}, \citenamefont {Mattila}, \citenamefont {H{\"a}m{\"a}l{\"a}inen},\ and\ \citenamefont {Sch{\"u}lke}}]{doring2004shake}%
  \BibitemOpen
  \bibfield  {author} {\bibinfo {author} {\bibfnamefont {G.}~\bibnamefont {D{\"o}ring}}, \bibinfo {author} {\bibfnamefont {C.}~\bibnamefont {Sternemann}}, \bibinfo {author} {\bibfnamefont {A.}~\bibnamefont {Kaprolat}}, \bibinfo {author} {\bibfnamefont {A.}~\bibnamefont {Mattila}}, \bibinfo {author} {\bibfnamefont {K.}~\bibnamefont {H{\"a}m{\"a}l{\"a}inen}},\ and\ \bibinfo {author} {\bibfnamefont {W.}~\bibnamefont {Sch{\"u}lke}},\ }\bibfield  {title} {\bibinfo {title} {\textit{Shake-up Valence Excitations in CuO by Resonant Inelastic x-Ray Scattering}},\ }\href@noop {} {\bibfield  {journal} {\bibinfo  {journal} {Phys. Rev. B}\ }\textbf {\bibinfo {volume} {70}},\ \bibinfo {pages} {085115} (\bibinfo {year} {2004})}\BibitemShut {NoStop}%
\end{thebibliography}%

\end{document}